\documentclass[journal, 12pt, draftclsnofoot, onecolumn]{IEEEtran}

\newcommand{\be}{\begin{equation}}
\newcommand{\ee}{\end{equation}}
\newcommand{\ba}{\begin{array}}
\newcommand{\ea}{\end{array}}
\newcommand{\bea}{\begin{eqnarray}}
\newcommand{\eea}{\end{eqnarray}}

\newcommand{\vbar}{\raisebox{.17ex}{\rule{.04em}{1.35ex}}}
\newcommand{\vbarind}{\raisebox{.01ex}{\rule{.04em}{1.1ex}}}
\newcommand{\R}{\ifmmode {\rm I}\hspace{-.2em}{\rm R} \else ${\rm I}\hspace{-.2em}{\rm R}$ \fi}
\newcommand{\T}{\ifmmode {\rm I}\hspace{-.2em}{\rm T} \else ${\rm I}\hspace{-.2em}{\rm T}$ \fi}
\newcommand{\N}{\ifmmode {\rm I}\hspace{-.2em}{\rm N} \else \mbox{${\rm I}\hspace{-.2em}{\rm N}$} \fi}
\newcommand{\B}{\ifmmode {\rm I}\hspace{-.2em}{\rm B} \else \mbox{${\rm I}\hspace{-.2em}{\rm B}$} \fi}
\newcommand{\Hil}{\ifmmode {\rm I}\hspace{-.2em}{\rm H} \else \mbox{${\rm I}\hspace{-.2em}{\rm H}$} \fi}
\newcommand{\C}{\ifmmode \hspace{.2em}\vbar\hspace{-.31em}{\rm C} \else \mbox{$\hspace{.2em}\vbar\hspace{-.31em}{\rm C}$} \fi}
\newcommand{\Cind}{\ifmmode \hspace{.2em}\vbarind\hspace{-.25em}{\rm C} \else \mbox{$\hspace{.2em}\vbarind\hspace{-.25em}{\rm C}$} \fi}
\newcommand{\Q}{\ifmmode \hspace{.2em}\vbar\hspace{-.31em}{\rm Q} \else \mbox{$\hspace{.2em}\vbar\hspace{-.31em}{\rm Q}$} \fi}
\newcommand{\Z}{\ifmmode {\rm Z}\hspace{-.28em}{\rm Z} \else ${\rm Z}\hspace{-.28em}{\rm Z}$ \fi}

\usepackage{graphicx}
\usepackage{subfig}
\usepackage{amsmath}
\usepackage{caption}
\usepackage{epsfig}
\usepackage{float}
\usepackage{lipsum}
\usepackage{stfloats}
\usepackage{array}
\usepackage{amssymb}
\usepackage{cite}
\usepackage{url}     
\usepackage{amsthm}
\usepackage{algorithm2e}
\usepackage{algorithmic}
\usepackage{multirow}
\usepackage{fancyh}
\usepackage{upgreek}
\usepackage{dsfont}
\usepackage{gensymb}
\normalsize

\newcommand{\Xr}{\mathsf{X}^{\scriptscriptstyle \rm r}_0}
\newcommand{\Xo}{X_0}
\newcommand{\Xk}{X_k}
\newcommand{\xo}{x_0}
\newcommand{\xk}{x_k}
\newcommand{\Yo}{X^{\scriptscriptstyle \rm c}_0}
\newcommand{\Yk}{X^{\scriptscriptstyle \rm c}_k}
\newcommand{\yo}{y_0}
\newcommand{\yk}{y_k}
\newcommand{\zo}{z_0}
\newcommand{\zk}{z_k}

\newcommand{\xik}{x_{i,k}}
\newcommand{\yik}{y_{i,k}}
\newcommand{\zik}{z_{i,k}}

\newcommand{\ro}{r_0}
\newcommand{\rk}{r_k}
\newcommand{\rio}{r_{i,0}}
\newcommand{\rik}{r_{i,k}}

\newcommand{\Yj}{X^{\scriptscriptstyle \rm c}_j}

\newcommand{\xr}{\mathsf{x}^{\scriptscriptstyle \rm r}_0}
\newcommand{\yr}{\mathsf{y}^{\scriptscriptstyle \rm r}_0}
\newcommand{\zr}{\mathsf{z}^{\scriptscriptstyle \rm r}_0}

\newcommand{\Xio}{X_{i,0}}
\newcommand{\Xik}{X_{i,k}}

\newcommand{\Yik}{X^{\scriptscriptstyle \rm c}_{i,k}}

\newcommand{\xcj}{x^{\scriptscriptstyle \rm c}_j}
\newcommand{\ycj}{y^{\scriptscriptstyle \rm c}_j}

\newcommand{\usrDia}{\mathcal{D}}

\newcommand{\beamW}{{\Omega}}

\newcommand{\incangik}{\theta_{i,k}}

\newcommand{\betaoo}{\beta_{0}}
\newcommand{\betaio}{\beta_{i,0}}
\newcommand{\betaik}{\beta_{i,k}}
\newcommand{\betaok}{\beta_{k}}

\newcommand{\rowvec}[1]{
  \begin{matrix}[#1]\end{matrix}%
}

\newcommand{\bGamma}{\boldsymbol{\Gamma}}
\newcommand{\bp}{\boldsymbol{p}}

\newcommand{\msfr}{\mathsf{r}_{\rm \scriptscriptstyle w}}

\newcommand{\mathX}{\mathbfcal{X}_0}

\newcommand{\novar}{\sigma_{\scriptscriptstyle \rm N}^2}

\newcommand{\imunit}{\mathsf{j}}

\newcommand{\Gtk}{\mathcal{G}_k^{\rm \scriptscriptstyle t}}
\newcommand{\Gtik}{\mathcal{G}_{i,k}^{\rm \scriptscriptstyle t}}
\newcommand{\Grk}{\mathcal{G}_k^{\rm \scriptscriptstyle r}}
\newcommand{\Grik}{\mathcal{G}_{i,k}^{\rm \scriptscriptstyle r}}

\newcommand{\Gto}{\mathcal{G}_0^{\rm \scriptscriptstyle t}}
\newcommand{\Gtio}{\mathcal{G}_{i,0}^{\rm \scriptscriptstyle t}}
\newcommand{\Gro}{\mathcal{G}_0^{\rm \scriptscriptstyle r}}
\newcommand{\Grio}{\mathcal{G}_{i,0}^{\rm \scriptscriptstyle r}}

\newcommand{\txorientEk}{\varsigma^{\rm \scriptscriptstyle e}_k}
\newcommand{\txorientAk}{\varsigma^{\rm \scriptscriptstyle a}_k}

\newcommand{\txorientEik}{\varsigma^{\rm \scriptscriptstyle e}_{i,k}}
\newcommand{\txorientAik}{\varsigma^{\rm \scriptscriptstyle a}_{i,k}}

\newcommand{\RXbeamE}{\uppsi^{\rm \scriptscriptstyle e}_0}
\newcommand{\RXbeamA}{\uppsi^{\rm \scriptscriptstyle a}_0}
\newcommand{\TXbeamEk}{\psi^{\rm \scriptscriptstyle e}_k}
\newcommand{\TXbeamAk}{\psi^{\rm \scriptscriptstyle a}_k}
\newcommand{\TXbeamEik}{\psi^{\rm \scriptscriptstyle e}_{i,k}}
\newcommand{\TXbeamAik}{\psi^{\rm \scriptscriptstyle a}_{i,k}}

\newcommand{\TXbeamEo}{\psi^{\rm \scriptscriptstyle e}_0}
\newcommand{\TXbeamAo}{\psi^{\rm \scriptscriptstyle a}_0}

\newcommand{\usrHt}{h_{\scriptscriptstyle \rm u}}

\newcommand{\Zup}{\mathcal{H}_{\scriptscriptstyle \rm up}}
\newcommand{\Zdw}{\mathcal{H}_{\scriptscriptstyle \rm dw}}

\newcommand{\ak}{a_k}
\newcommand{\bk}{b_k}
\newcommand{\aktilde}{\tilde{a}}
\newcommand{\bktilde}{\tilde{b}}

\newcommand{\patG}{\mathcal{G}}

\newcommand{\Gamlow}{\Gamma_{\scriptscriptstyle \rm low}}
\newcommand{\Gamhigh}{\Gamma_{\scriptscriptstyle \rm high}}

\DeclareMathAlphabet\mathbfcal{OMS}{cmsy}{b}{n}

\theoremstyle{definition}

\newtheorem{exmp}{Example}

\allowdisplaybreaks
\begin{document}
\bstctlcite{IEEEexample:BSTcontrol}

\title{Enclosed mmWave Wearable Networks: Feasibility and Performance}
\author{Geordie~George,~\IEEEmembership{Student Member,~IEEE}, Kiran~Venugopal,~\IEEEmembership{Student Member,~IEEE}, Angel~Lozano,~\IEEEmembership{Fellow,~IEEE} and~Robert~W.~Heath,~Jr.,~\IEEEmembership{Fellow,~IEEE}
\thanks{G. George and A. Lozano are with the Department of Information and Communication Technologies, Universitat Pompeu Fabra (UPF), 08018 Barcelona, Spain. E-mail: \{geordie.george, angel.lozano\}@upf.edu. 
        
K. Venugopal and R. W. Heath Jr. are with The University of Texas at Austin, Austin, TX 78704-0240. E-mail: \{kiranv, rheath\}@utexas.edu. 
      
This work was supported in part by the Intel/Verizon University Research Program ``5G: Transforming the Wireless User Experience'' and by Project TEC2015-66228-P (MINECO/FEDER, UE), as well as by the European Research Council under the H2020 Framework Programme/ERC grant agreement 694974. Parts of this paper were presented at the IEEE Int'l Workshop on Computational Advances in Multi-Sensor Adaptive Processing (CAMSAP'15) \cite{CAMSAP15}.}}

\maketitle

\begin{abstract}
This paper investigates the feasibility of mmWave frequencies for personal networks of wireless wearable devices in enclosed settings (e.g., commuter trains, subways, airplanes, airports, or offices).
At these frequencies, specular reflections off surfaces are expected to contribute 
intended signal power and, simultaneously, to aggravate the interference at the receivers. Meanwhile, blockages by obstacles and people---including the individuals wearing the devices---are expected to shield receivers from interference.
With the aid of stochastic geometry and random shape theory, we assess the interplay of surface reflections and blockages for dense deployments of wearable networks equipped with directional antenna arrays in relevant indoor settings.
\end{abstract}



\section{Introduction} 
\label{sec:introduction}
The expanding market for wearable computing devices (in short, \emph{wearables})  bespeaks of a tomorrow where the sight of people with multiple body-born wearables connected wirelessly might become commonplace \cite{Juniper13,6926659}.
The communication among wearables is expected to be highly proximal, in the form of small body area networks composed of very-short-range on-body links \cite{6739368}, with a wide range of bit rate requirements: from low-rate activity trackers to high-rate augmented-reality devices \cite{6844949,Oculus}. 
The presence of several wearable networks---one per person---in close vicinity creates a very high density of simultaneous wireless transmissions. 
While transmissions within each wearable network can be orthogonalized by means of coordination via a hub, interference from other wearable networks is very likely, as coordination across people may be unfeasible. 
Understanding the ensuing complex interference environment as well as the on-body wireless channel is crucial to assess the communication performance of such networks.

Operation at mmWave (millimeter wave) frequencies seems promising for wearable networks due to inherent characteristics of these frequencies, namely the availability of bandwidth (e.g., in the 60 GHz unlicensed band), the suitability for short-rage communication and dense spectral reuse, and the practicality of implementing directional antenna arrays within small devices \cite{4300986,Rapp-mmWave13,mmWaveBook-Ted}. 
MmWave communication for indoor applications is becoming a reality thanks to standards such as WirelessHD \cite{WirelessHD} and IEEE 802.11ad \cite{5421713}. These standards, or the proposed D2D (device-to-device) communication modes in mmWave-based 5G systems \cite{Boccardi-5G,What5G,7147834} could potentially be employed for wearable networks.
There is, therefore, interest in establishing the feasibility of deploying very dense mmWave wearable networks, chiefly in enclosed settings \cite{Alex-Kerstin15,7437379,KirHeath15,mmWaveD2D_Kiran15,7445132}.

At mmWave frequencies, signals exhibit reduced scattering and minimal diffraction around blocking obstacles, but strong specular reflections off surfaces \cite{Smulders1994_reflect, Fernandes1994, 588558, 995521, 5262304,5454109,6042312,mmWaveSurvey15}. As the blocking by obstacles---including people themselves---results in huge propagation losses \cite{HumanBlock13,Collonge2004,mmWaveSurvey15}, surface reflections are expected to play a major role in the performance of enclosed mmWave networks, by contributing additional signal and interference powers. 
By means of directional beamforming \cite{mmWaveBook-Ted, 7145934}, wearables can gather useful signal from intended directions while reducing some of the unwanted interference incoming from other directions.
 
In this paper, we investigate the impact of reflections and blockages on the fundamental performance limits of enclosed mmWave networks with emphasis on dense deployments and with wearables equipped with directional antennas. 
The propagation models in \cite{KirHeath15, mmWaveD2D_Kiran15, KirHeathAsilomar15} accounted for reflections in a coarse way, by fitting different pathloss parameters for the LOS (line-of-sight) and the NLOS (non-line-of-sight) links, rendering them more amenable to analysis. Differently, recognizing that the pathloss parameter values reported by different indoor measurements vary significantly \cite{5208265},
we set out to model reflections explicitly. The multipath propagation environment resulting from the surface reflections is modeled via geometric optics, surface reflectivity and free space pathloss, similar to the models in \cite{5454109,6042312}.
As for the blockages, we build on \cite{KirHeath15, mmWaveD2D_Kiran15}, where human body blockages in direct links were modeled explicitly but reflections were not, by incorporating the reflections off interior surfaces and accounting for blockages in both direct and reflected paths. Measurements reported in \cite{588558} have verified that the characteristics of mmWave indoor propagation that determine radio link performance are chiefly dictated by the reflections off the superstructure (i.e., walls, ceiling and floor) while the influence of details such as tables and cabinets is insignificant. Nonetheless, in the crowded scenarios of our interest, incorporating reflections off the human bodies \cite{BodyReflectionTed} might be a necessary follow-up to our work.

Based on the approach in \cite{BaiHeath-blockage14,6932503}, of applying stochastic geometry and random shape theory to analyze the building blockage effects in outdoor cellular networks, \cite{mmWaveD2D_Kiran15} devised a stochastic model for the body blockages in direct propagation paths. We expand this stochastic blockage model, incorporating blockages in the reflected paths as well, so as to obtain results without the need to exhaustively test whether each individual link is blocked.

Ultimately, we seek to understand whether reflections are beneficial or detrimental, and whether satisfactory performance is possible in relevant enclosed settings.
Considering the additional signal and interference contributions due to reflections, and the capability for directional beamforming, several examples of the performance of a reference transmitter-receiver pair in the network are provided, to answer questions such as:
\begin{itemize}
	\item How does the performance vary with surface reflectivity and signal blockages?
	\item Denser environments mean more sources of interference, but also more interference blockages. What is the net effect, with surface reflections accounted for?
	\item Is the performance limited by interference or by noise?
	\item How does the relative location of the reference transmitter-receiver pair affect the performance?
	\item In the absence of a strong direct signal path, do the reflections provide enough useful signal for satisfactory operation? If so, what range of beamforming gains are necessary? 
\end{itemize}

\section{Network Modeling}
\label{Net Mod}

Consider people within an enclosed space with reflective interior surfaces and no signal penetration from outside. Each individual wears multiple communication devices and the intended transmissions are always between devices on a same person. Those transmissions are assumed to be orthogonal as they can be coordinated via a hub. Therefore, interfering transmissions are always from wearables on different people.

We focus on a time-frequency  channel occupied by a reference transmitter-receiver pair on a reference person. There are $K$ other people on which the interfering transmitters reusing the same channel are located. Each person has one transmitter on the channel under consideration.

\subsection{Network Geometry}
\label{Net Geo}
\begin{figure}
	\centering
	\includegraphics [width=0.65\columnwidth]{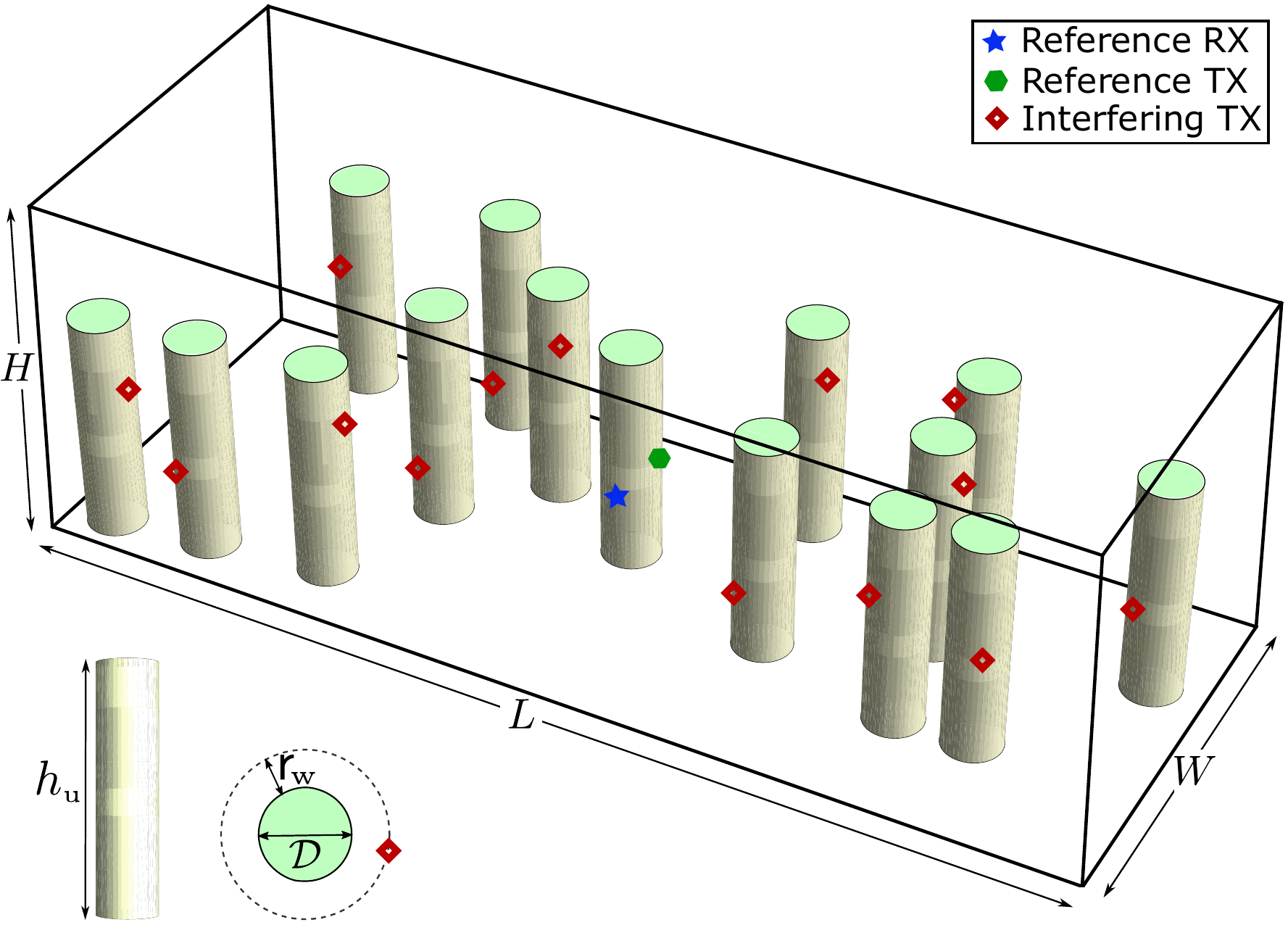}
	\caption{Cochannel wearable devices on people---modeled as circular cylinders of diameter $\usrDia$ and height $\usrHt$---within an enclosed space. Each wearable is located below height $\usrHt$ and at a horizontal distance $\msfr \geq 0$ from its body.}
	\label{Fig:figure0}
\end{figure}

We consider an enclosed space shaped as an $L \times W \times H$ cuboid (cf. Fig. \ref{Fig:figure0}) with people modeled as cylinders of diameter $\usrDia$, height $\usrHt < H$ and axis perpendicular to the floor. Each wearable is located below height $\usrHt$, at a perpendicular distance $\usrDia/2 + \msfr$ from the axis of its cylinder and with an azimuth orientation random in $[0,2\pi)$.
In effect, $\msfr \geq 0$ is the distance of each wearable from its body.
The reference receiver is located at $\Xr$ while the $K+1$ transmitters are located at $\{\Xk\}_{k=0}^K$, with $\Xo$ being the intended (reference) transmitter. 
With $L$ along the $x$-axis, $W$ along the $y$-axis, $H$ along the $z$-axis, and fixing the origin at the center of the enclosed space, let the coordinates of $\Xr$ and $\Xk$ be respectively $(\xr,\yr,\zr)$ and $(\xk,\yk,\zk)$, while the distance between $\Xr$ and $\Xk$ is $\rk = \|\Xk-\Xr\|$.

\subsection{Surface Reflections}

\begin{figure}
	\centering
	\includegraphics [width=0.65\columnwidth]{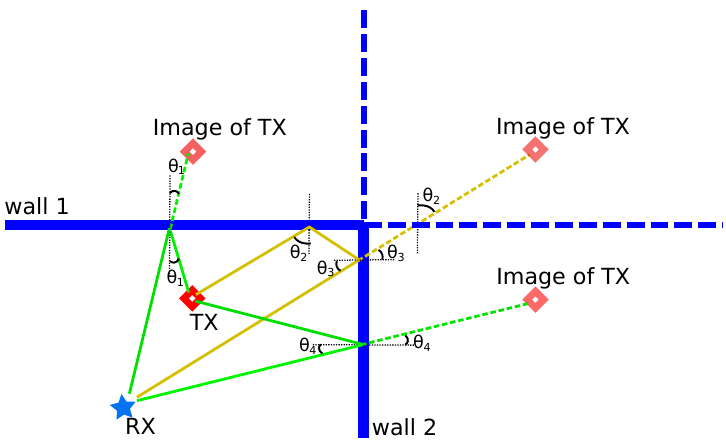}
	\caption{Reflected links from a transmitter to a receiver, off two walls. There are two first order reflections and one second order reflection.}
	\label{Fig:figure1}
\end{figure}
The transmission from $\Xk$ reaches $\Xr$ via a direct propagation link and via reflections off the surfaces. To model these reflections, we need the lengths of the reflected links as well as the angles of incidence and the ensuing reflection coefficients.

\subsubsection{Geometry of the Reflections}

Adding extra (phantom) transmitters at the mirror image locations across each surface (cf. Fig. \ref{Fig:figure1}) facilitates the reflection modeling \cite{Images-Reinaldo-1992}.
In this paper, we consider only first-order reflections, i.e., single bounces off each surface.\footnote{We have numerically verified that higher-order reflections, which can be incorporated by placing phantom transmitters at the corresponding image locations (cf. Fig. \ref{Fig:figure1}), have a minor effect on the results.} 
From each transmitter $\Xk$ there are six such reflections reaching $\Xr$, which are incorporated by adding six phantom transmitters. The four walls are indexed with $i=1,\ldots,4$, the ceiling with $i=5$, and the floor with $i=6$. For $i=1,\ldots,6$, the images of $\Xk$ are located at $\Xik$, the corresponding angles of incidence are $\incangik$, and the reflected link distances are $\rik = \|\Xik - \Xr\|$. The coordinates of the image locations and the angles of incidence can be easily obtained as functions of the coordinates of $\Xk$ and $\Xr$, as detailed in Appendix \ref{App1}. 

Note that the links emanating from $\{\Xio\}_{i=1}^6$ correspond to the reflections of the intended transmission from $\Xo$. While the intended transmission has a direct \emph{on-body} link and six reflected \emph{off-body} links, all the interfering links (both direct and reflected) are off-body.

\subsubsection{Reflection Coefficient}
\label{reflection coefficient}

The reflectivity of a surface depends on the properties of the material, the angle of incidence, and the polarization of the incident wave.
We apply the model in \cite{Sato95}, which provides
reflection coefficients $\Gamma_\perp$ and $\Gamma_\parallel$ for a homogeneous dielectric plate with a smooth surface, thickness $\Delta$ and complex refractive index $n$. 

These coefficients are
\begin{align} 
\Gamma_\ell  = \frac{1 - e^{-\imunit \, 2 \,  \delta}}{1 - \gamma_\ell^2 \, e^{- \imunit \, 2 \, \delta}} \, \gamma_\ell \qquad \ell \in \{\perp,\parallel\}
\end{align}
with
\begin{align} \nonumber
\delta &= \frac{2 \pi \Delta}{\lambda} \sqrt{n^2 - \sin^2 \theta} \\
\gamma_\perp &= \frac{\cos \theta-\sqrt{n^2 -\sin^2 \theta}}{\cos \theta+\sqrt{n^2 -\sin^2 \theta}}\\
\nonumber 
\gamma_\parallel &= \frac{n \cos \theta-\sqrt{n^2 -\sin^2 \theta}}{n \cos \theta+\sqrt{n^2 -\sin^2 \theta}} 
\end{align}
where $\lambda$ is the wavelength.
The coefficients $\Gamma_\perp$ and $\Gamma_\parallel$ relate the reflected and incident electric fields when the polarization is respectively perpendicular and parallel to the plane of incidence (defined as the plane that contains the incident and reflected rays and the surface normal).

As per the Rayleigh criterion \cite{Smulders1994_reflect}, the shorter wavelength of mmWave signals renders surfaces rougher than at microwave frequencies. While many indoor surfaces might still be deemed smooth, the effect of roughness---when it is significant---features as an extra loss factor in the reflection coefficient, which can either be modeled based on the standard deviation of surface roughness or be incorporated implicitly when the reflectivity is measured \cite{Smulders1994_reflect, Fernandes1994,588558}. Importantly, at mmWave frequencies, the ensuing diffuse scattering off rough surfaces has been observed to not contribute significantly to the total received power, such that the propagation remains effectively specular \cite{Smulders1994_reflect, Fernandes1994, 588558}.

The following example presents two extreme reflectivity settings, low reflectivity ($\Gamlow$) and high reflectivity ($\Gamhigh$), which will be employed throughout the paper to gauge the impact of surface reflectivity on the performance.

\begin{exmp}
\begin{figure}[!tbp]
  \centering
  \subfloat[Magnitude]{\includegraphics[width=0.4\linewidth]{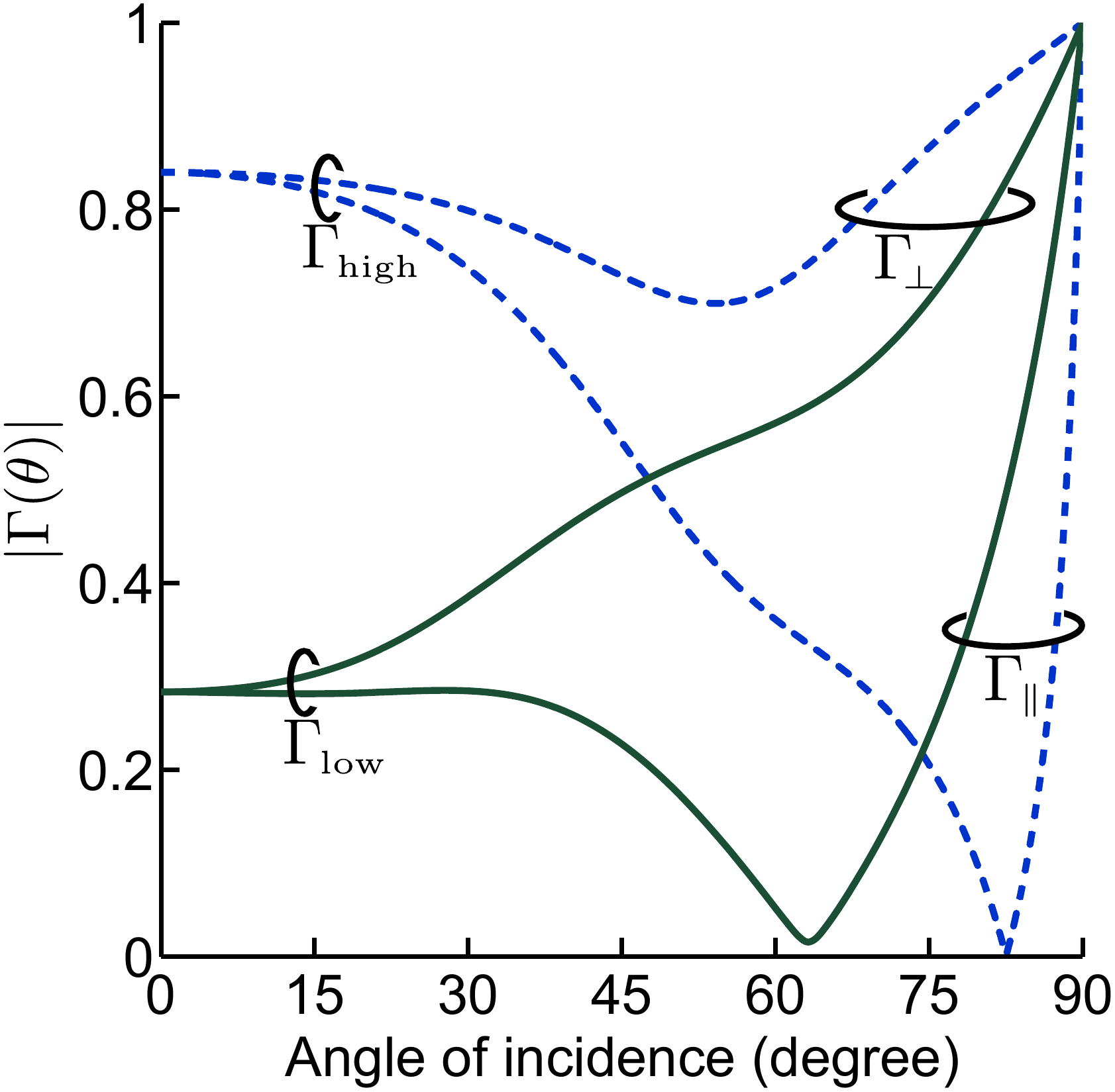}\label{fig:f1}}
\qquad
  \subfloat[Phase]{\includegraphics[width=0.4\linewidth]{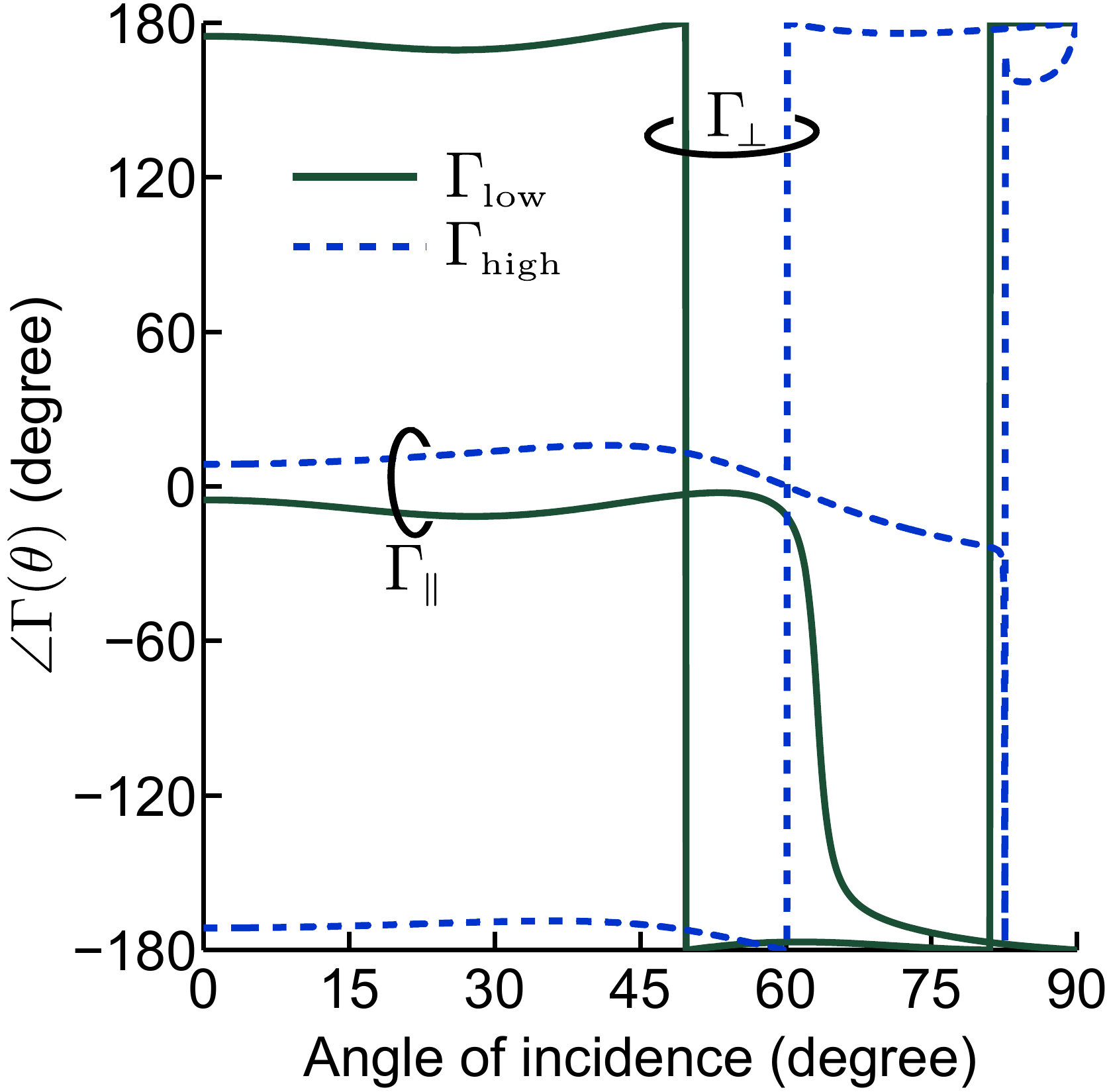}\label{fig:f2}}
  \caption{Low ($\Gamlow$) and high ($\Gamhigh$) reflection coefficients as functions of the angle of incidence.}
  \label{Fig:refcoef}
\end{figure}
Fig. \ref{Fig:refcoef} depicts the reflection coefficients at 60 GHz (i.e., $\lambda = 5 \, {\rm mm}$) obtained with the following values of $\Delta$ and $n$:
\begin{enumerate}
\item Low reflectivity ($\Gamlow$): $\Delta = 8.8 \, {\rm mm}$ and $n = 7.62 - \imunit \, 0.02$
\item High reflectivity ($\Gamhigh$): $\Delta = 14.2 \, {\rm mm}$ and $n = 1.85 - \imunit \, 0.086$
\end{enumerate}
\end{exmp}


To account for the random orientation of the wearables, the polarization of the electric field is regarded as random.
The direction of the electric field vector---perpendicular to the direction of propagation by definition---in each propagation path from $X_k$ is abstracted by an angle (cf. Appendix \ref{App11}), which is denoted by $\alpha_k$ for the direct path and by $\alpha_{i,k}$ for the $i$th reflected path.
Let us define the corresponding polarization unit vectors 
\begin{align}\label{eq:bpk}
	\bp_k &= \rowvec{\cos \alpha_{k} & \sin \alpha_{k}}^{\rm T}\\
	\bp_{i,k} &= \rowvec{\cos \alpha_{i,k} & \sin \alpha_{i,k}}^{\rm T}\label{eq:bpik}
\end{align}
which shall come handy later, in expressing the propagation model in Section \ref{Prop Model}.
Upon reflection on a surface, the field is projected with reference to the plane of incidence, the appropriate reflection coefficient ($\Gamma_\parallel$ or $\Gamma_\perp$) is applied to each projected component, and the field is subsequently reconstructed \cite{5454109,6042312}.
Specifically, assuming that all surfaces have the same $\Delta$ and $n$, the horizontally and vertically polarized components of each transmission are respectively subject to $\Gamma_\parallel$ and $\Gamma_\perp$ when bouncing off walls ($i=1,\ldots,4$), and vice versa when bouncing off ceiling or floor ($i=5,6$). Correspondingly, we define the reflection coefficient matrix
\begin{align}\label{eq:RefCeofMat}
\bGamma_{i,k} = \left\{ \begin{array}{l l}
   \bGamma(\incangik) \quad &  \quad i = 1, \ldots, 4 \\
   \bGamma^{\rm T}(\incangik) \quad &  \quad i = 5,6
 \end{array} \right.
\end{align}
with $\bGamma(\theta) = \left[\begin{smallmatrix}
       \Gamma_\parallel(\theta) & 0            \\[0.2em]
       0 & \Gamma_\perp(\theta)          
\end{smallmatrix}\right]$, also to be applied in the expressions in Section \ref{Prop Model}. 
Readers interested in further details on how reflections affect polarization are referred to \cite{529140}.

\subsection{Body Blockages}
\label{body blockages}

The links (both direct and reflected) among wearables  can get blocked by people's bodies. 
Since the reflections are modeled explicitly, and the penetration losses at mmWave frequencies are very high---typically in excess of $40$ dB---we assume that no signal traverses such blockages. The blocking of the direct link from $\Xk$ is indicated by a binary variable $\betaok$, which equals $1$ if unblocked and $0$ if blocked. Likewise, the blocking of the link from $\Xk$ reflected off the $i$th surface is indicated by another binary variable  $\betaik$. 

\subsubsection{Interference Path}
\label{blockage interference}

The direct interference path between $\Xk$ and $\Xr$ is blocked if it cuts through any of the cylinders. Such blockages include self-body blockages \cite{BaiHeath15}, i.e., the link between $\Xk$ and $\Xr$ can get blocked by the people on which transmitter or receiver are located.
Since everybody has the same height $\usrHt$ and the wearables are located below that height, the blockages in the direct interfering links are independent of the heights of the wearable locations. Therefore, the blockages can be determined from the projections of the wearable locations and the cylinders onto the horizontal plane, which we denote by $\mathX$, that contains $\Xr$. Noting that the projections of the cylinders are circles on $\mathX$, we denote the projection of $\Xk$ by $\Xk'$. Then, as in \cite{KirHeath15, mmWaveD2D_Kiran15, CAMSAP15}, the blockages can be determined by checking whether the direct path between $\Xk'$ and $\Xr$ intersects any of the circles. Applied to the corresponding phantom transmitters, this blockage model further extends to the reflected links off the four walls and an algorithm for determining such blockages is given in Appendix \ref{App2}.

%

\begin{figure}
	\centering
	\includegraphics [width=0.7\columnwidth]{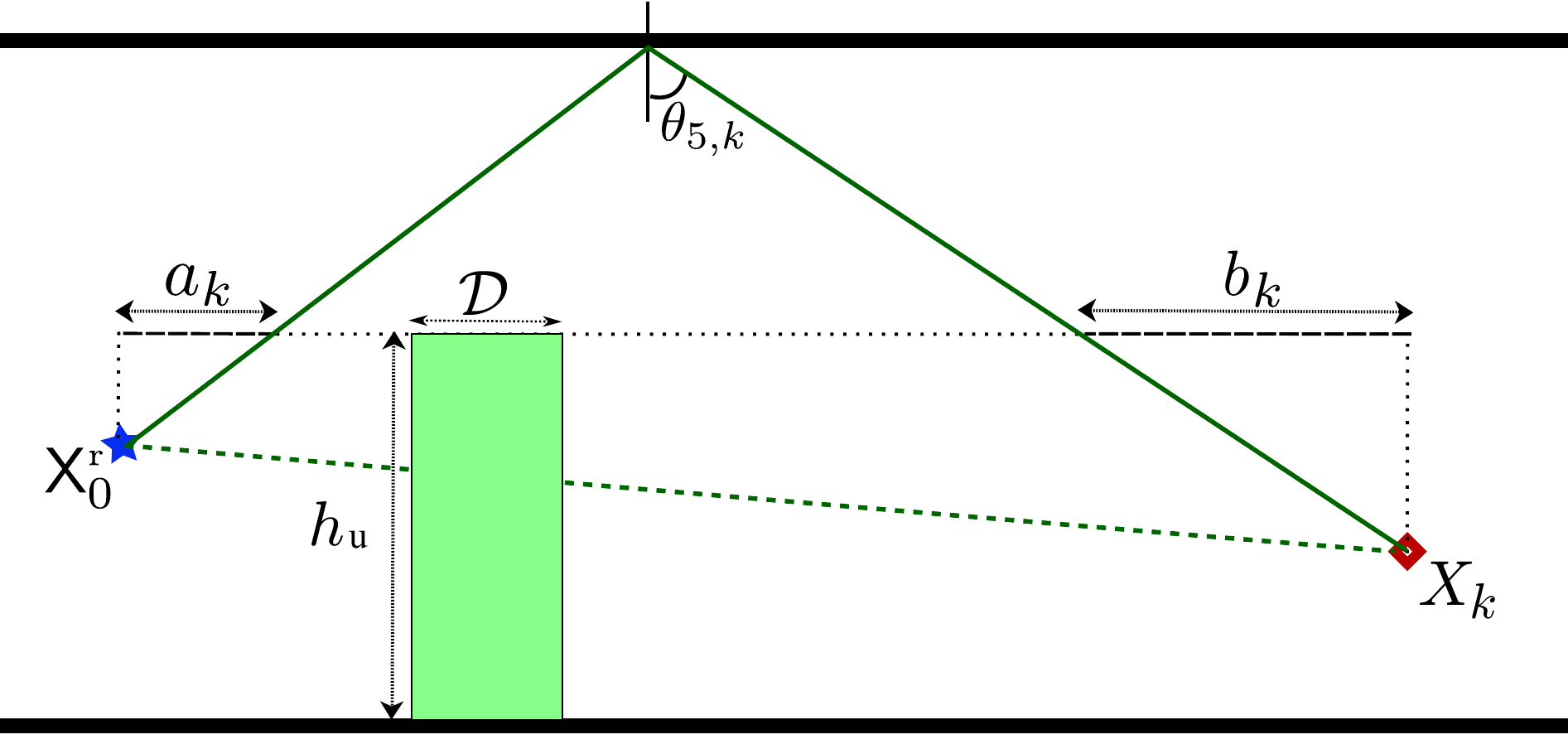}
	\caption{Ceiling-reflected link from transmitter $\Xk$ to receiver $\Xr$. Bodies within the horizontal distances $\ak$ and $\bk$ block the ceiling reflection.}
	\label{Fig:figure2}
\end{figure}
Given the gap between the ceiling and people's heads, a ceiling reflection is blocked only if someone is close enough to the transmitter or receiver, specifically closer than
\be
\ak = (\usrHt - H/2 - \zr) \, \tan \theta_{5,k}
\ee
from the receiver at $\Xr$ or closer than 
\be
\bk = (\usrHt - H/2 - \zk) \, \tan \theta_{5,k}
\ee
from the transmitter at $\Xk'$ (cf. Fig. \ref{Fig:figure2}).
Further details on how to determine the ceiling blockages are given in Appendix \ref{App2}.

Since bodies touch the floor, each reflection off the floor gets blocked only if the corresponding direct path is blocked, i.e., $\beta_{6,k} = \betaok$.

\subsubsection{Intended Signal Path}
\label{intended signal paths}

The foregoing blockage model for the wall reflections is applicable to all the off-body links, including the wall-reflected paths of the intended transmission from $\Xo$. The coefficients $\betaio$, for $i=1,\ldots,4$, are determined via the algorithm in Appendix \ref{App2}.

As for the on-body intended link, since it is between wearables on the same individual, it should have an independent blockage/shadowing model.
In the absence of a good model for on-body shadowing, we consider specific values for $\betaoo$ in the range $[0,1]$. Two values of special interest are: (\emph{unblocked on-body link}) $\betaoo=1$ and (\emph{blocked on-body link}) $\betaoo=0$. When the on-body link is blocked, transmission from $\Xo$ reaches $\Xr$ only via the reflected links. 

The intended signal reflections off the ceiling and the floor are assumed to be unblocked ($\beta_{5,0} = \beta_{6,0}=1$). A possible future refinement for $\beta_{5,0}$ and $\beta_{6,0}$ would be to model them based on specific wearable applications.

\section{Antenna Arrays}
\label{Antenna Arrays}

\begin{figure}
	\centering
	\includegraphics [width=0.45\columnwidth]{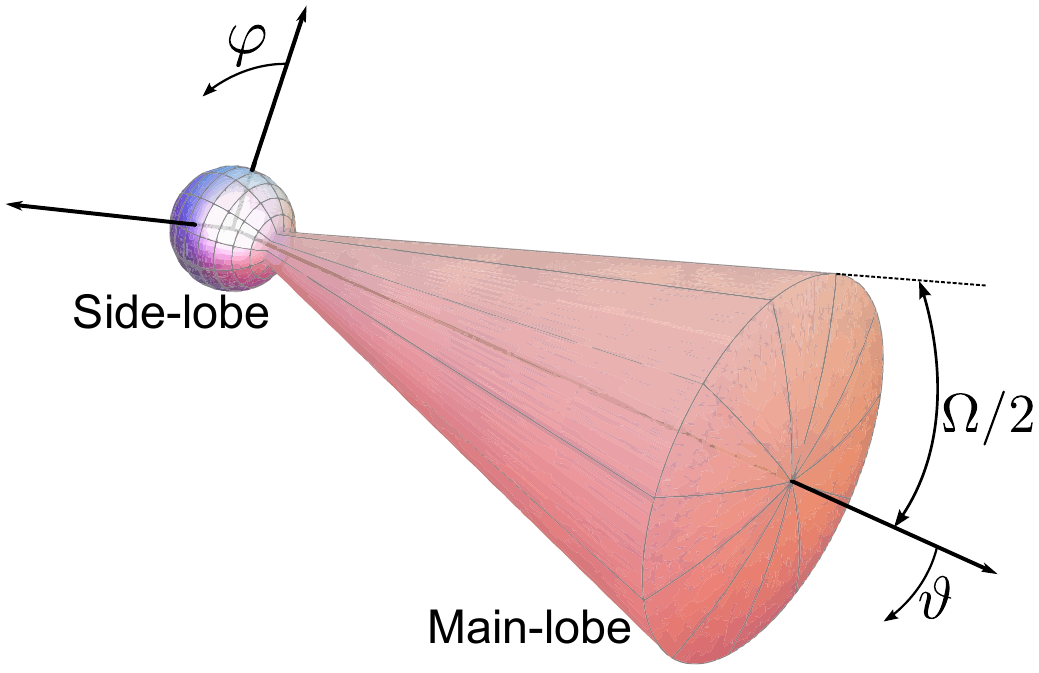}
	\caption{Antenna gain pattern, rotationally symmetric about the broadside direction.}
	\label{Fig:ant}
\end{figure}

At mmWave frequencies, devices are expected to incorporate polarization-diverse antennas so as to circumvent polarization mismatch losses; this is in fact rather critical given the limited scattering, and thus the reduced depolarization, experienced at these frequencies \cite{7147840,Lawrence150322,lawrence2016tri_orthogonal}. We therefore postulate that polarization diversity is in place, providing
immunity from polarization mismatch losses. The antennas most likely to be featured by wearables are patches, dual-polarized versions of which have been implemented lately \cite{6569608,7081095}.
An array of such patches produces a beam that can be steered by physically titling the array or else through beamforming coefficients.
We model this beam as having rotational symmetry
with two defining parameters: the main-lobe gain, $G$, and its beamwidth, $\beamW$ (cf. Fig. \ref{Fig:ant}).
Expressed as a function of $\vartheta$ (off the beam axis) and $\varphi$ (on the plane perpendicular to the beam axis), 
\begin{align}\label{eq:antpat}
\patG(\vartheta,\varphi) = \left\{ \begin{array}{l l}
 G \quad &  0 \leq \vartheta \leq \beamW/2, \, 0 \leq \varphi < 2\pi \\
 g \quad &  \beamW/2 < \vartheta \leq \pi, \, 0\leq \varphi < 2\pi
 \end{array} \right.
\end{align}
with $g$ the side-lobe gain and with 
\be
\int_{0}^{2\pi} \int_{0}^{\pi} \patG(\vartheta,\varphi) \, \frac{\sin \vartheta}{ 4\pi} \, {\rm d}\vartheta \, {\rm d}\varphi = 1
\ee
to ensure that the total radiated power is preserved \cite{balanis:2012}.
While simple, (\ref{eq:antpat}) captures the key features very  effectively, approximating well the pattern of an $N$-element uniform planar square array (UPA), i.e., a $(\sqrt{N}\times\sqrt{N})$ UPA with main-lobe gain $G = N$ and beamwidth $\beamW = \sqrt{3/N}$, and consequently with
\begin{align}\label{eq:sidelobegain}
g = N + (1-N) \, \sec^2\left[\sqrt{3/N}/4\right].
\end{align}

\begin{exmp}
Listed in Table \ref{tab:Antenna Params} are the antenna array parameters for different values of $N$ that are used in later examples in this paper. Note that $N = 1$ corresponds to omnidirectional antennas.
\begin{table}
	\caption{Array Settings}
	\label{tab:Antenna Params}
	\vspace{-0.15in}
	\begin{center}
		\begin{tabular}{ |c|c|c|c|}
			\hline
			$N$ & $G$ (dBi) & $g$ (dBi) & $\Omega$    \\ \hline\hline
			1 & 0 & 0  &     \\ \hline
			4 & 6 & -0.68 & 49.6$^\circ$ \\ \hline
			\end{tabular}
			\begin{tabular}{ |c|c|c|c|}
				\hline
				$N$ & $G$ (dBi) & $g$ (dBi) & $\Omega$    \\ \hline\hline
				9 & 9.54 & -0.80 & 33$^\circ$ \\ \hline
				16 & 12 & -0.85 & 24.6$^\circ$ \\ \hline
			\end{tabular}
		\end{center} 
		\vspace{-0.25in}
\end{table}
\end{exmp}

To incorporate the transmit and receive antenna gains to the propagation from transmitter $\Xk$ to the reference receiver $\Xr$ via the direct and the reflected paths, we introduce (\emph{i}) the transmit antenna gains $\Gtk$ and $\Gtik$, respectively from $\Xk$ and $\Xik$ in the direction of $\Xr$, and (\emph{ii}) the receive antenna gains $\Grk$ and $\Grik$, respectively in the direction of $\Xk$ and $\Xik$. 

By virtue of the rotationally symmetric pattern, the receive antenna gains can be obtained by evaluating (\ref{eq:antpat}) for
$\vartheta$ equal to the angle between the main-lobe direction of $\Xr$ and the orientation of the transmitter relative to $\Xr$.
Similarly, the transmit antenna gains can be determined by evaluating (\ref{eq:antpat}) for $\vartheta$ equal to the angle between the main-lobe direction of transmitter and the orientation of $\Xr$ relative to the transmitter.
We specify the main-lobe directions of the wearables with the azimuth and the elevation angles with respect to the $x$, $y$ and $z$ axes as defined in Section \ref{Net Geo}: the main-lobe of $\Xr$ is directed at azimuth $\RXbeamA$ and elevation $\RXbeamE$, while $\Xk$ has its main-lobe directed at azimuth $\TXbeamAk$ and elevation $\TXbeamEk$. Further details on the determination of the gain parameters and the main-lobe directions of the phantom transmitters are given in Appendix \ref{App3}.

\section{Propagation Model}
\label{Prop Model}

All transmissions have (fixed) power $P$ and each wearable is assumed to have its main-lobe oriented towards its intended signal link (direct on-body or reflected off-body). 


\subsection{Intended Signal}

The intended transmission from $\Xo$ is received at $\Xr$ with power
\begin{align}\label{eq:signal_rec}
\mathsf{P}_0 &=  P \, \left(\frac{\lambda}{4\pi}\right)^2 \, \bigg\|\betaoo \, \frac{\sqrt{\Gro \, \Gto}}{\ro} \, \bp_0 + \sum_{i=1}^{6} \betaio \, \frac{\sqrt{\Grio \, \Gtio}}{\rio}\,  e^{-\imunit \, \Delta\phi_{i,0}} \, \bGamma_{i,0} \, \bp_{i,0} \bigg\| ^2
\end{align}
where 
$\Delta\phi_{i,0} = 2\pi(\rio-\ro)/\lambda$ is the phase difference between the direct link and the $i$th reflected link.
Given the locations $\Xr$ and $\Xo$, their respective main-lobe directions ($\{\RXbeamA, \RXbeamE\}$ and $\{\TXbeamAo, \TXbeamEo\}$) and subsequently all the antenna gain parameters in (\ref{eq:signal_rec}) become determined (cf. Appendix \ref{App3}).


Recall
that the vectors $\bp_0$ and $\bp_{i,0}$,
respectively given in (\ref{eq:bpk}) and (\ref{eq:bpik}), 
abstract the polarization, 
while the reflection coefficient matrix $\bGamma_{i,0}$ is determined via (\ref{eq:RefCeofMat}). 
Also recall, from Section \ref{intended signal paths}, that the blocking coefficient $\betaoo$ is specified independently of the relative locations of the wearables while $\{\betaio\}_{i=1}^4$ are determined as per the algorithm in Appendix \ref{App2} and $\beta_{5,0} = \beta_{6,0}=1$.

\subsection{Interference} 

The transmission from $\Xk$, for $k=1,\ldots,K$, is received at $\Xr$ with power
\begin{align}\label{eq:power_received}
\mathsf{P}_k &=  P \, \left(\frac{\lambda}{4\pi}\right)^2 \, \bigg\|\betaok \, \frac{\sqrt{\Grk \, \Gtk}}{\rk} \, \bp_k + \sum_{i=1}^{6} \betaik \, \frac{\sqrt{\Grik \, \Gtik}}{\rik}\,  e^{-\imunit \, \Delta\phi_{i,k}} \, \bGamma_{i,k} \, \bp_{i,k} \bigg\| ^2
\end{align}
where $\Delta\phi_{i,k} = 2\pi(\rik-\rk)/\lambda$ is the phase difference between the direct and the $i$th reflected links.
Given the locations $\Xk$ and $\Xr$, and the receiver main-lobe direction $\{\RXbeamA, \RXbeamE\}$, the receive antenna gains in (\ref{eq:power_received}) are determined, as detailed in Appendix \ref{App3}. The transmit antenna gains depend on the main-lobe direction of $\Xk$, which is assumed to be uniformly distributed, i.e., $\TXbeamAk$ is uniform in $[0,2\pi)$, and  $\TXbeamEk$ is distributed with PDF (probability density function) \cite{SpherePointP}
\be
f_{\TXbeamEk}(\nu) = \frac{\sin \nu}{2} \quad 0 \leq \nu \leq \pi.
\ee
Again, recall that $\boldsymbol{p}_{k}$, $\boldsymbol{p}_{i,k}$ and $\bGamma_{i,k}$ are respectively as given in (\ref{eq:bpk}), (\ref{eq:bpik}) and (\ref{eq:RefCeofMat}), while ${\betaok}$ and $\{\betaik\}_{i=0}^5$ are determined as per Appendix \ref{App2} and $\beta_{6,k}=\betaok$.


\section{SINR}

The
SINR (signal-to-interference-plus-noise ratio) at the reference receiver is
\begin{align}\label{eq:rho}
\mathsf{SINR} = \frac{\mathsf{P}_0}{\novar + \sum_{k=1}^{K} \mathsf{P}_k}
\end{align}
where $\novar = F_{\rm N} N_0 B$ is the AWGN power, with
$F_{\rm N}$ the receiver noise figure, $N_0$ the noise power spectral density and $B$ the bandwidth.

For a specific network geometry (i.e., given the positions of people and wearables) and specific orientations
and polarization angles, $\mathsf{SINR}$ in (\ref{eq:rho}) becomes determined.
A randomized network geometry and a distribution for the orientation of people
and the polarization angles
induce a distribution for $\mathsf{SINR}$.
In the next section, we introduce a random network geometry model---specified by the distribution of wearables and blockages within the enclosure---and stochastically model the concomitant propagation parameters so as to simplify the computation of the SINR distribution.

\section{Stochastic Modeling}
\label{Stoch Model}

Stochastic geometry analyses of unbounded wireless networks, modeled via appropriate point processes, are usually conducted from the perspective of the typical receiver---randomly chosen and held fixed---over all possible network geometries. Differently, for the finite-size space under consideration, we evaluate the performance of a reference link with given receiver location $\Xr$, when the interfering transmitters $\{X_k\}_{k=1}^K$ and the blockages (people) are distributed randomly.
Such an evaluation is arguably more informative than a complete averaging over all possible locations of the reference link, as the performance does depend on such location: a link in the center of the space, for instance, will generally perform differently than a link near one of the corners.


\subsection{Random Network Geometry}
\label{user distribution}

The reference individual, represented by the corresponding circle on $\mathX$, has its center $\usrDia/2+\msfr$ away from $\Xr$ with a uniformly random orientation in $[0,2\pi)$. The corresponding transmitter $\Xo$ is another wearable on the reference individual at a distance $\ro$ away from $\Xr$. The coordinates of $\Xo$ are
\begin{align}\nonumber
\xo &= \xr + \ro \sin\varsigma^{\scriptscriptstyle e}_0 \cos\varsigma^{\scriptscriptstyle a}_0 \\
\label{eq:\Xo}
\yo &= \yr + \ro \sin\varsigma^{\scriptscriptstyle e}_0 \sin\varsigma^{\scriptscriptstyle a}_0 \\
\nonumber
\zo &= \zr + \ro \cos\varsigma^{\scriptscriptstyle e}_0
\end{align}
where $\varsigma^{\scriptscriptstyle a}_0 \in [0,2\pi)$ and $\varsigma^{\scriptscriptstyle e}_0$ has the PDF
\begin{align}
f_{\varsigma^{\scriptscriptstyle e}_0}(\nu) = \frac{\sin \nu}{2} \qquad 0 \leq \nu \leq \pi
\end{align}
meaning that the direction of $\Xo$ from $\Xr$ is uniformly distributed.
\begin{figure}
	\centering
	\includegraphics [width=0.65\columnwidth]{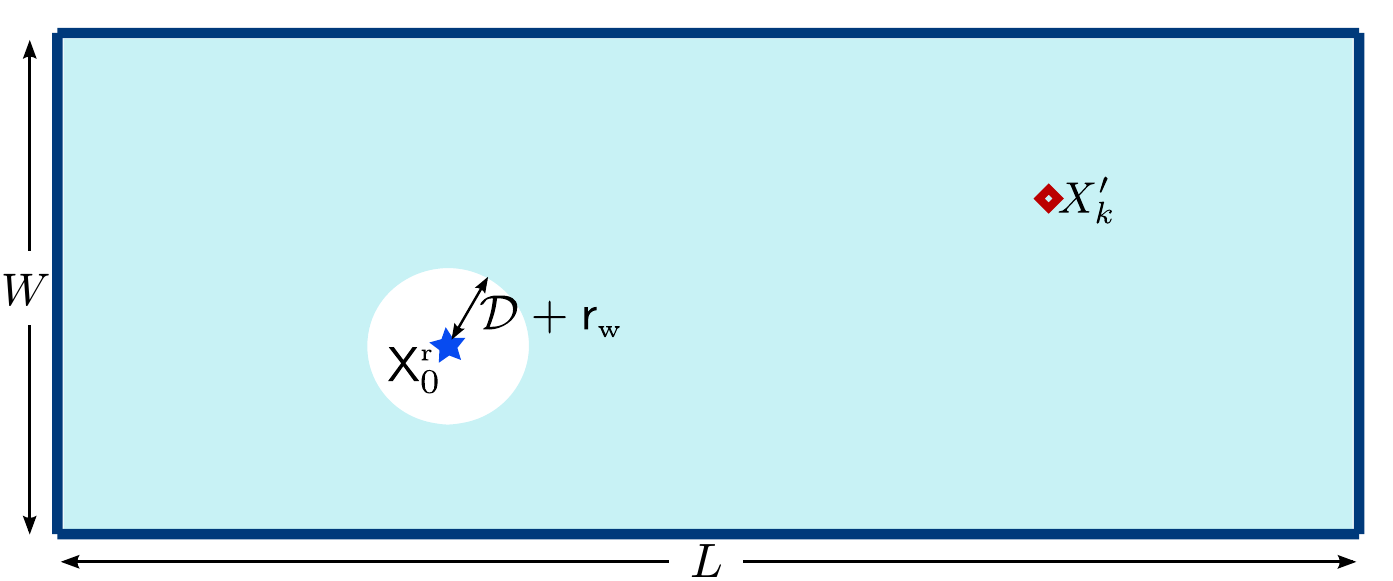}
	\caption{Horizontal plane $\mathX$ containing the reference receiver $\Xr$. Projections of the interfering wearables $\{X_k'\}_{k=1}^K$ are distributed independently and uniformly over the shaded area.}
	\label{Fig:horizontalplane}
\end{figure}

The interfering wearables, $\{X_k\}_{k=1}^K$, are distributed independently as follows. Each projection $\Xk'$, specified by the coordinates $\xk$ and $\yk$, is uniformly distributed on $\mathX$ excluding a circular region centered at $\Xr$ with radius $\usrDia + \msfr$ (cf. Fig. \ref{Fig:horizontalplane}).\footnote{This small exclusion region, consistent with the impossibility of two people occupying the same physical space, is introduced to avoid circles overlapping with the reference receiver and thereby blocking all the links. Interfering wearables overlapped by circles are considered blocked (cf. Appendix \ref{App2}).} 
As the center of $\mathX$ is at $(0,0,\zr)$, the joint PDF of $\xk$ and $\yk$ is \begin{align}\label{eq:JPDFxy}
f_{\xk,\yk}(x,y) = \frac{1}{LW - \pi (\usrDia+\msfr)^2}   \qquad\quad
 \begin{aligned}
 & |x|<L/2, \;  |y|<W/2  \\
 & (x-\xr)^2+(y-\yr)^2 > (\usrDia+\msfr)^2
 \end{aligned}
\end{align}
while $\zk$, which specifies the $k$th interfering wearable's height, varies independently and uniformly in $[\Zup,\Zdw]$, with $\Zup < \usrHt - H/2$ and $\Zdw > -H/2$.

Similarly, the individual wearing the interferer at $\Xk$, represented by the corresponding circle on $\mathX$, has its center $\usrDia/2+\msfr$ away from the projection $\Xk'$, at an angle uniform in $[0,2\pi)$. The distribution of such circle center on $\mathX$, while not exactly uniform, can be approximated by the uniform distribution in (\ref{eq:JPDFxy})---tighter as $\usrDia/2+\msfr$ gets smaller compared to $L$ and $W$---so as to derive (cf. Appendix \ref{App4}) the approximate blockage probabilities presented next. The accuracy of these approximations 
will be validated, for practically relevant settings, in Examples \ref{av block prob} and \ref{SINR CDF}.

\subsection{Stochastic Blockage Model}
\label{blockage probability}

We now set out to devise, by means of random shape theory  \cite{BaiHeath-blockage14,mmWaveD2D_Kiran15}, a stochastic alternative to the algorithm presented in Appendix \ref{App2} to determine blockages.


\subsubsection{Direct Interference Paths}

For the direct interference path from a given transmitter $\Xk$ to $\Xr$, considering the potential blockages by the $K-1$ other people and the potential self-body blocking by the link's own individuals (the reference one and the one wearing $\Xk$), the probability of blockage satisfies (cf. Appendix \ref{App4}) 
\begin{align} \label{eq:probblk} 
\mathbb{P}[\betaok = 0] 
\approx 1 - \left( 1 -\frac{ \rk' \, \usrDia  - \mathcal{A}}{LW - \pi (\usrDia+\msfr)^2}\right)^{K-1}  \left(1 - \frac{\arcsin\frac{\usrDia}{2 \, \msfr+\usrDia}}{\pi} \right)^2
\end{align}
where $\rk' = \|\Xk' - \Xr\|$ and 
\begin{align}
\mathcal{A} =   (\usrDia+\msfr)^2  \, \arcsin\frac{\usrDia/2}{\usrDia+\msfr} + \frac{\usrDia \, (\usrDia+\msfr)}{2} \, \cos \left(\arcsin\frac{\usrDia/2}{\usrDia + \msfr}\right) - \pi \usrDia^2/8.
\end{align}
Then, $\betaok$ for each $\Xk$ is a Bernoulli random variable with $\mathbb{P}[\betaok = 0]$ given by (\ref{eq:probblk}).

The self-body blockage probability, obtained by evaluating (\ref{eq:probblk}) with $K=1$, increases with shrinking $\msfr$ for given $\usrDia$, while being independent of the transmitter location. 
When $\msfr = 0$, a given link can get blocked by its own individuals with probability $3/4$. Quite naturally, blocking by other people occurs with higher probability for longer links and higher densities.

\subsubsection{Wall-Reflected Interference Paths}

As shown in Appendix \ref{App4}, the probability of blockage of the reflected interfering links off the walls, i.e., the links from $\{\Xik\}_{i=1}^4$, can also be approximated by (\ref{eq:probblk}). Thus, for $i=1,\ldots,4$, $\betaik$ is Bernoulli with 
\begin{align}
\mathbb{P}[\betaik = 0] \approx 1 - \left( 1 -\frac{ \rik' \, \usrDia - \mathcal{A}}{LW - \pi (\usrDia+\msfr)^2}\right)^{K-1}  \left(1 - \frac{\arcsin\frac{\usrDia}{2 \, \msfr+\usrDia}}{\pi} \right)^2
\end{align}
where $\rik' = \|\Xik' - \Xr\|$.

\subsubsection{Ceiling-Reflected Interference Paths}

To characterize the blockage probability of the ceiling reflection from $\Xk$, we express 
\be
\beta_{5,k} = \upbeta_{5,k}^{\scriptscriptstyle {\rm sb}0} \, \upbeta_{5,k}^{\scriptscriptstyle {\rm sb}k} \, \upbeta_{5,k}^{\scriptscriptstyle \rm ob}
\ee
where the independent Bernoulli variables $\upbeta_{5,k}^{\scriptscriptstyle {\rm sb}0}$, $\upbeta_{5,k}^{\scriptscriptstyle {\rm sb}k}$ and $\upbeta_{5,k}^{\scriptscriptstyle \rm ob}$ respectively capture the self-body blocking by the reference individual, self-body blocking by the person wearing $\Xk$, and blocking by the $K-1$ other people. Recalling $\ak = (\usrHt - H/2 - \zr) \, \tan \theta_{5,k}$ and $ \bk = (\usrHt - H/2 - \zk) \, \tan \theta_{5,k}$ from Section \ref{body blockages}, with a modicum of trigonometry (cf. Appendix \ref{App4}) we obtain 
\begin{align} \label{eq:celilSB0}
\mathbb{P}[\upbeta_{5,k}^{\scriptscriptstyle {\rm sb}0} = 0] = \left\{ \begin{array}{l l}
  \frac{\arcsin\frac{\usrDia}{2 \, \msfr+\usrDia}}{\pi}  \quad &  \quad \ak \geq \sqrt{\msfr (\usrDia+\msfr)} \\
  \frac{\arccos\frac{\ak^2 +  \msfr \usrDia + \msfr^2}{\ak (2 \, \msfr+\usrDia)}}{\pi}  \quad &  \quad \sqrt{\msfr (\usrDia+\msfr)} > \ak \geq \msfr \\ 
 0 \quad &  \quad \ak < \msfr
 \end{array} \right.
\end{align}
\begin{align} \label{eq:celilSBk}
\mathbb{P}[\upbeta_{5,k}^{\scriptscriptstyle {\rm sb}k} = 0] = \left\{ \begin{array}{l l}
\frac{\arcsin\frac{\usrDia}{2 \, \msfr+\usrDia}}{\pi}  \quad &  \quad \bk \geq \sqrt{\msfr (\usrDia+\msfr)} \\
  \frac{\arccos\frac{\bk^2 +  \msfr \usrDia + \msfr^2}{\bk (2 \, \msfr+\usrDia)}}{\pi}  \quad &  \quad \sqrt{\msfr (\usrDia+\msfr)} > \bk \geq \msfr \\ 
 0 \quad &  \quad \bk < \msfr
 \end{array} \right.
\end{align}
\begin{align}\label{eq:eq:celilOB}
\mathbb{P}[\upbeta_{5,k}^{\scriptscriptstyle \rm ob} = 0] \approx 1 - \left( 1 -\frac{ (\ak + \bk) \, \usrDia - \mathcal{A}+ \pi \, \usrDia^2/2 }{LW - \pi (\usrDia+\msfr)^2}\right)^{K-1}.  
\end{align}
Note that, unlike for direct path and wall reflections, the blockage probability of the ceiling reflection does depend on the wearable heights (via $\zk$ for the transmitter and $\zr$ for the receiver). 

Blocking of a ceiling reflection implies that the corresponding direct path is also blocked, i.e, $\beta_{5,k}$ and $\betaok$ are dependent and
\begin{align}\label{eq:directblockprob}
\mathbb{P}[\betaok = 0] &= \mathbb{P}[\betaok = 0 | \beta_{5,k} = 0] \, \mathbb{P}[\beta_{5,k} = 0] + \mathbb{P}[\betaok = 0 | \beta_{5,k} = 1] \, \mathbb{P}[\beta_{5,k} = 1] \\
&=\mathbb{P}[\beta_{5,k} = 0] + \mathbb{P}[\betaok = 0 | \beta_{5,k} = 1] \, \mathbb{P}[\beta_{5,k} = 1].
\end{align}
This dependence can be captured in the stochastic model by introducing an auxiliary random variable, as explained in Appendix \ref{App4}.
\subsubsection{Wall-Reflected Signal Paths}

As for the intended signal reflections off the walls, i.e., the links from $\{\Xio\}_{i=1}^4$, only the reference individual can effect self-body blockage on them while the other $K$ people can potentially intersect the links. Then, as argued in Appendix \ref{App4}, the probability of blockage for the link from $\{\Xio\}_{i=1}^4$ satisfies
 \begin{align}\label{eq:block prob sig ref}
\mathbb{P}[\betaio=0] &\approx  1 - \left( 1 -\frac{ \rio' \, \usrDia - \mathcal{A}}{2 \, LW - 2 \, \pi (\usrDia+\msfr)^2 }\right)^{K} \left(1 - \frac{\arcsin\frac{\ro'}{2 \, \msfr+\usrDia} + \arcsin\frac{\usrDia}{2 \, \msfr+\usrDia}}{\pi} \right)
\end{align}
with $\rio' = \|\Xio'-\Xr\|$ and $\ro' = \|\Xo'-\Xr\|$. For $i=1,\ldots,4$, the coefficient $\betaio$ is Bernoulli with $\mathbb{P}[\betaio=0]$ given by (\ref{eq:block prob sig ref}).

Note that the self-body blockage of the wall-reflected signal paths (by the reference individual) depends on the intended transmitter location $\Xo$, in addition to $\msfr$ and $\usrDia$. 


\begin{table}
	\caption{Settings}
	\label{tab:NetSet}
	\vspace{-0.15in}
	\begin{center}
		\begin{tabular}{ |c|c|}
			\hline
			Parameter & Value  \\ \hline\hline
			$L \times W \times H$ & 20 m $\times$ 4 m $\times$ 2.5 m \\ \hline
			$\ro$ & 25 cm \\ \hline
		    $\lambda$ & 5 mm \\ \hline
		\end{tabular}
		\begin{tabular}{ |c|c|}
			\hline
		 	 Parameter & Value  \\ \hline\hline
			 $\usrDia$ & 50 cm \\ \hline
			 $\usrHt$ & 175 cm \\ \hline
			 $\Zup$, $\Zdw$ & 25 cm, -75 cm \\ \hline
		\end{tabular}
		\begin{tabular}{ |c|c|}
			\hline
			Parameter & Value  \\ \hline\hline
   		$P$ & 0 dBm \\ \hline
			$F_{\rm N}$ &  9 dB \\ \hline
			$N_0$  & -174 dBm/Hz \\ \hline
		\end{tabular}
	\end{center}
\end{table}


The following example validates, for the settings in Table \ref{tab:NetSet}, the blockage probabilities that we have established throughout this section.

\begin{exmp}\label{av block prob}

\begin{figure}
	\centering
	\includegraphics [width=0.75\columnwidth]{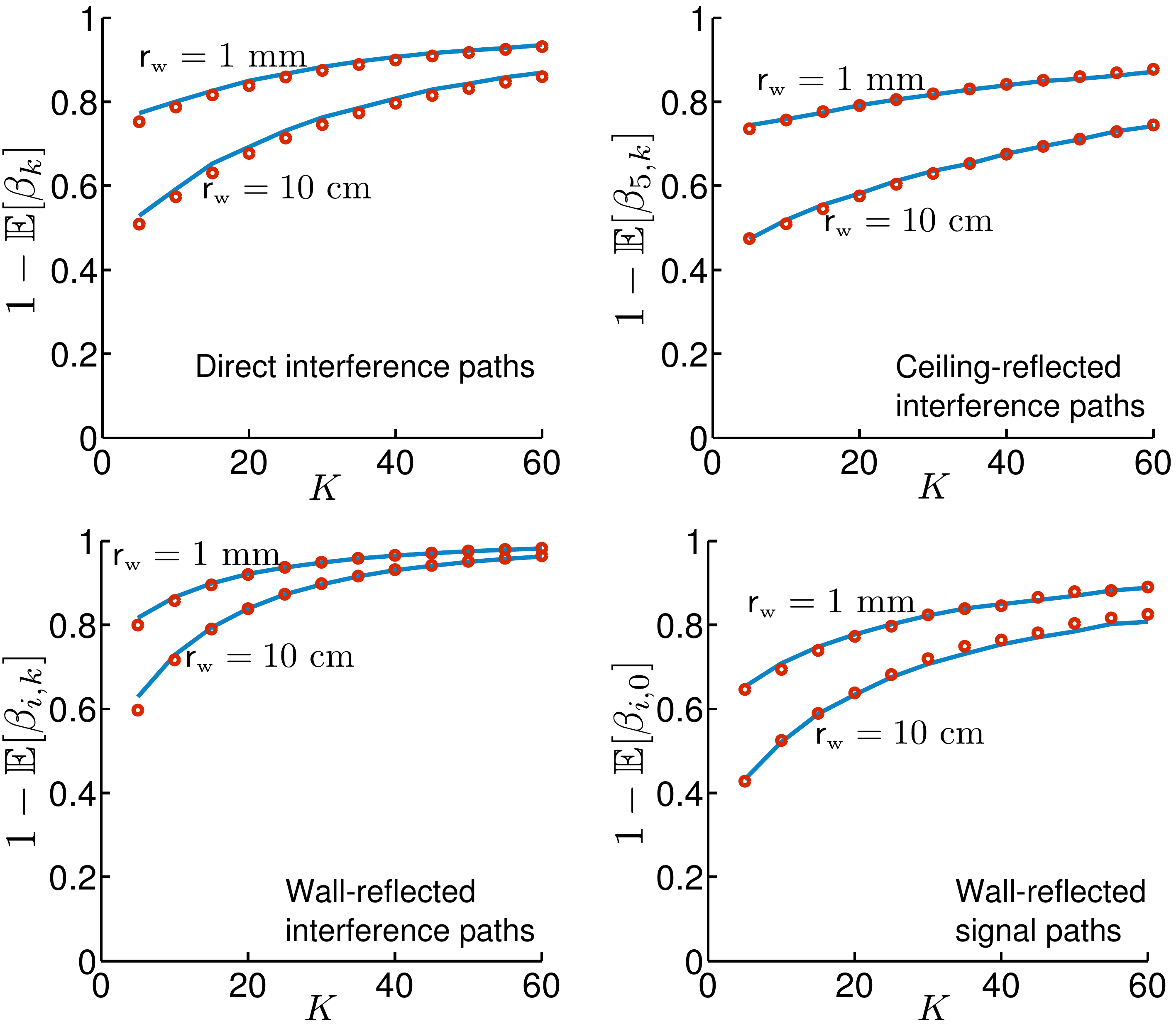}
	\caption{Average blockage probabilities of the direct interference path and the reflected paths.}
	\label{Fig:valid1}
\end{figure}

Consider a reference receiver $\Xr$ at the center of the enclosed space.
Depicted with markers in Fig. \ref{Fig:valid1} are the average blockage probabilities, computed via the expectations of the location-dependent blockage random variables over the locations distributed as per Section \ref{user distribution}, for different values of $K$ and $\msfr$. 
Their exact counterparts, obtained via Monte-Carlo (by establishing each individual blockage deterministically as detailed in Appendix \ref{App2}), are the solid curves plotted alongside.
\end{exmp}

Very good matches in support of our stochastic model are observed. As can be seen, the blockage probability increases significantly with decreasing $\msfr$ (self-body blockage) and with increasing densities (other-body blockage).

From the marginal distributions established for $\{ \betaok \}$ and $\{ \betaik \}$, a stochastic blockage model can be constructed by regarding these variables as independent, functions only of their respective transmitter locations.
This ignores potential dependences across links due to common blockages and related reflections, and thus some validation of whether significant such dependences do exist is needed before we can confidently apply the model.
This validation is provided in Example \ref{SINR CDF}.

\subsection{Stochastic Model for the Antenna Gains}
\label{stoch mod ant G}

Since the main-lobe directions of the interfering transmitters are distributed independently and uniformly, the receiver $\Xr$ is in the main-lobe of $\Xk$ with probability
\begin{align}
p_{\scriptscriptstyle \rm t} &= \int_{0}^{2 \pi}\int_{0}^{\frac{\sqrt{3/N}}{2}} \frac{\sin \vartheta}{4 \, \pi} \, {\rm d}\vartheta \,  {\rm d}\varphi \\
&= \sin^2\left(\sqrt{3/N}/4\right).
\end{align}
Then, the transmit antenna gain $\Gtk$, for $k = 1,\ldots,K$, satisfies
\begin{align}\label{eq:PMFGtk}
\Gtk =  \left\{ \begin{array}{l l}
G & \quad \mathrm{with} \; \mathrm{prob.} \quad p_{\scriptscriptstyle \rm t}   \\
g & \quad  \mathrm{with} \; \mathrm{prob.} \quad 1 - p_{\scriptscriptstyle \rm t} 
 \end{array} \right.
\end{align}
where $G = N$ and $g$ is as in (\ref{eq:sidelobegain}). The transmit antenna gains for the reflected links, $\Gtik$, have the same distribution. The dependence between $\Gtk$ and $\Gtik$ is ignored, with the accuracy of this assumption validated in Example \ref{SE CDFs}.

Recall, from Section \ref{Antenna Arrays} and Appendix \ref{App3}, that all the receive antenna gains $\{\Grk\}$ and $\{\Grik\}$ are functions of the main-lobe direction of $\Xr$. Consequently, they are mutually dependent for $k = 1,\ldots,K$, unlike the transmit antenna gains.
In this paper, we intend to study the network performance under specific cases of the receiver main-lobe direction (cf. Section \ref{Imp Ant Arr}).
Nonetheless, over a uniform distribution of the receiver main-lobe direction, each $\Grk$ and $\Grik$ would also abide by (\ref{eq:PMFGtk}). 

\section{Impact of Reflections and Blockages}
\label{results}

This section provides examples, for the settings in Table \ref{tab:NetSet}, 
to test the accuracy of the stochastic blockage model proposed in Section \ref{blockage probability} and to gauge the impact of reflections and blockages on the communication performance with isotropic antennas ($N = 1$). The results presented hereafter are obtained for two specific locations for the reference receiver $\Xr$:
\begin{itemize}
\item \emph{Center} of the space, whereby $\Xr$ is at the origin
\item \emph{Corner} of the space, whereby $(\xr, \yr, \zr) = (8.5 \,\,\text{m}, 1.5 \,\,\text{m}, 0.25 \,\,\text{m})$.
\end{itemize}
Also, recall the two reflectivity settings ($\Gamlow$ and $\Gamhigh$) from Fig. \ref{Fig:refcoef}. 
Unless otherwise specified, the bandwidth $B$, which determines the noise power $\novar$, is set to $B = 1$ GHz.
\begin{figure}[!tbp]
	\centering
	\subfloat[Unblocked on-body link ($\betaoo = 1$)]{\includegraphics[width=0.4\linewidth]{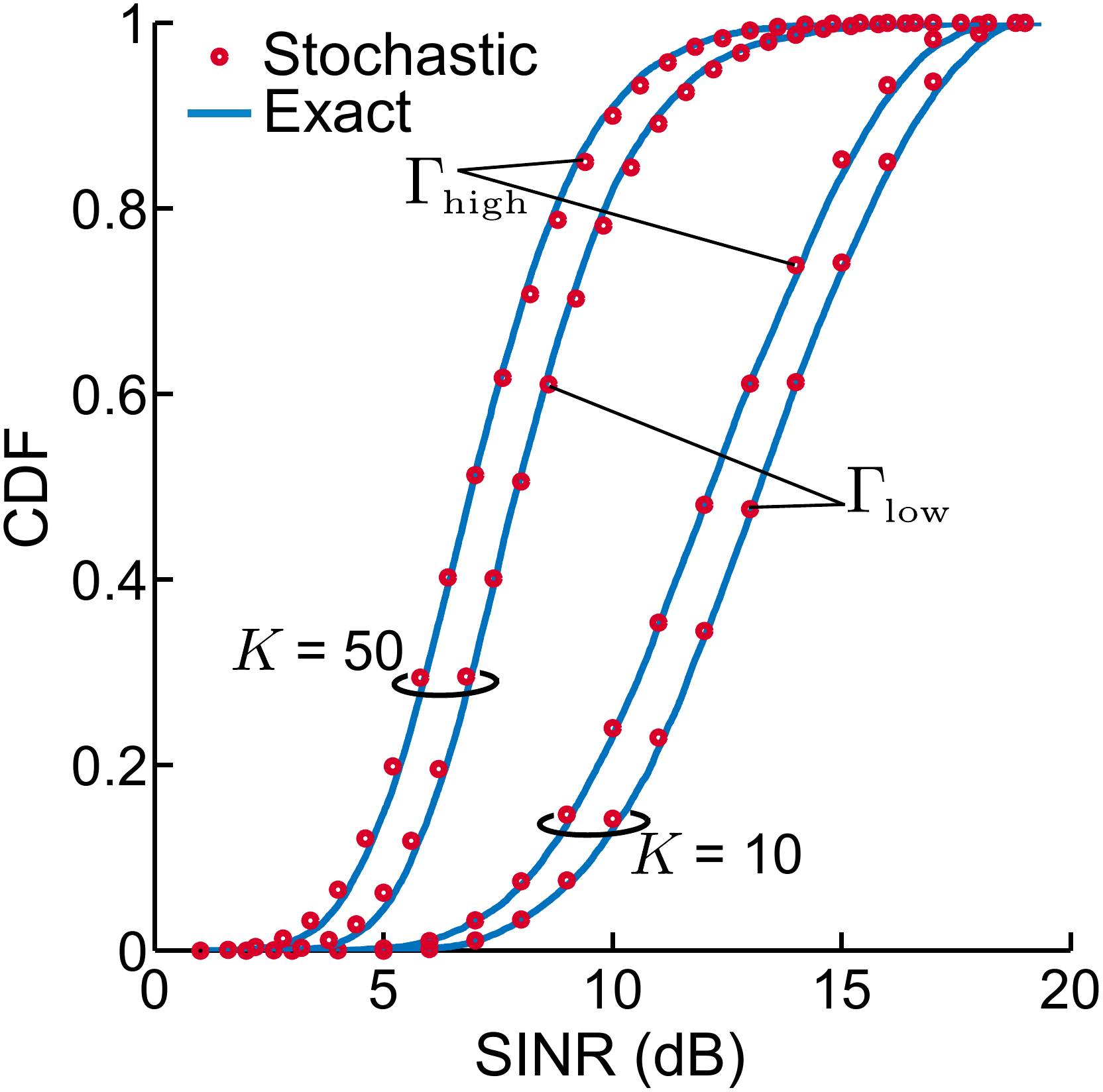}}
	\qquad
	\subfloat[Blocked on-body link ($\betaoo = 0$)]{\includegraphics[width=0.4\linewidth]{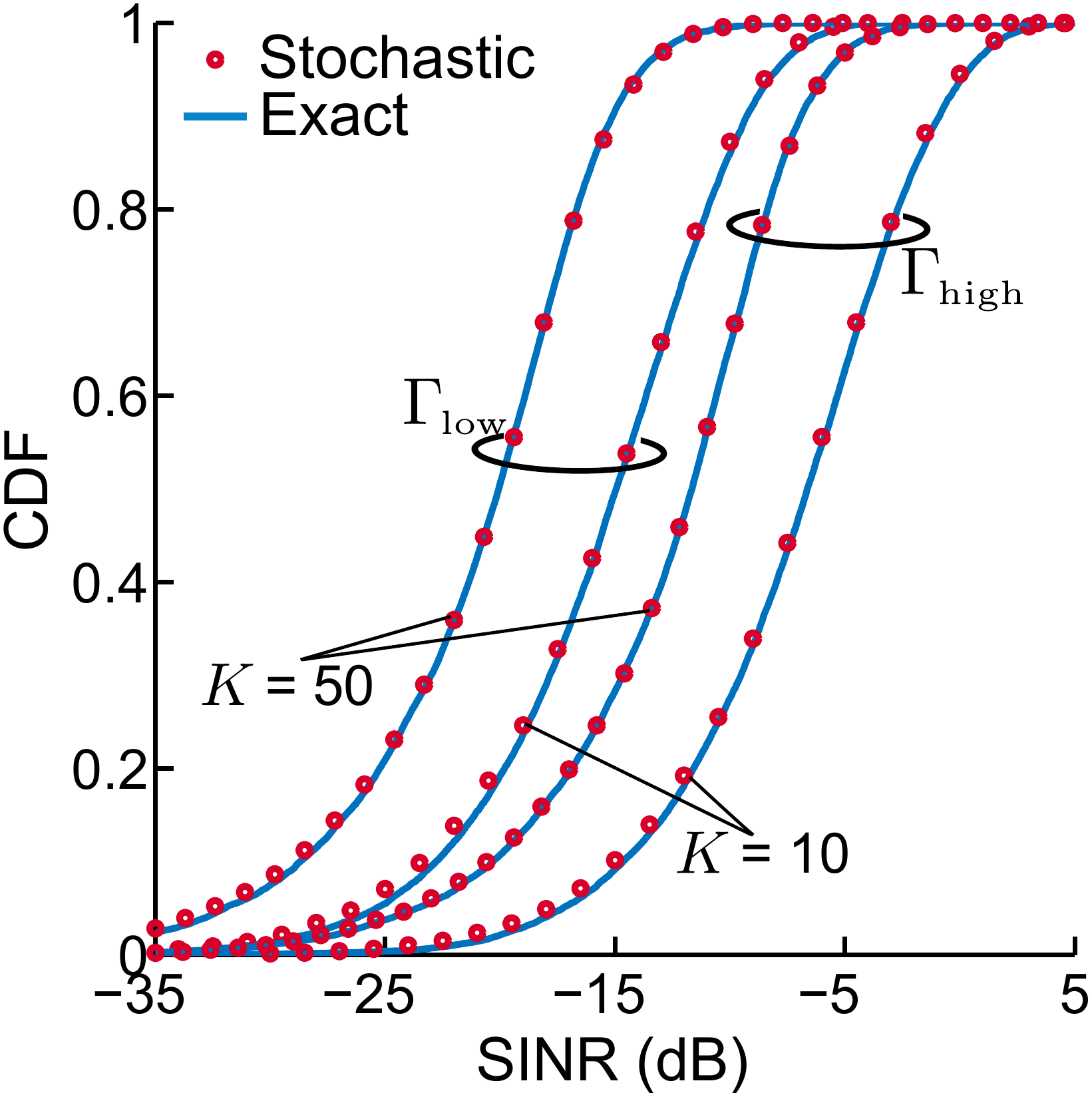}}
	\caption{CDFs of SINR with the reference receiver located at the center, $\msfr = 10$ cm and isotropic antennas ($N = 1$).}
	\label{Fig:figure7}
\end{figure}
\begin{exmp}
\label{SINR CDF}
For a reference receiver located at the center, the CDFs of $\mathsf{SINR}$ are plotted in Fig. \ref{Fig:figure7}, with $\msfr = 10$ cm and $N = 1$. The CDFs obtained by independently realizing $\{ \betaok \}$ and $\{ \betaik \}$ via the probabilities given in Section \ref{blockage probability}, ignoring their dependences, are contrasted against the exact ones obtained by establishing each individual blockage deterministically as detailed in Appendix \ref{App2}.
\end{exmp}

Very good agreements are observed,
supporting the stochastic model under these settings and indicating that the distribution of $\mathsf{SINR}$ computed over all possible geometries is not sensitive to potential dependences caused by common blockages and related reflections. 

Given the minimal fading at mmWave frequencies and under the premise of Gaussian signaling, the spectral efficiency can be obtained from $\mathsf{SINR}$ as 
$
C (\mathsf{SINR}) = \log_2(1+\mathsf{SINR}) ,
$
which can then be spatially averaged over the distribution of $\mathsf{SINR}$ (dictated by all possible locations of people and wearables, orientations and polarization angles) to obtain the average performance for a given reference receiver location
\be\label{eq:avgSE}
\bar{C} = \mathbb{E}[\log_2(1+\mathsf{SINR})].
\ee

\begin{figure}[!tbp]
	\centering
	\subfloat[Receiver at center and $\msfr = 10$ cm.]{\includegraphics[width=0.4\linewidth]{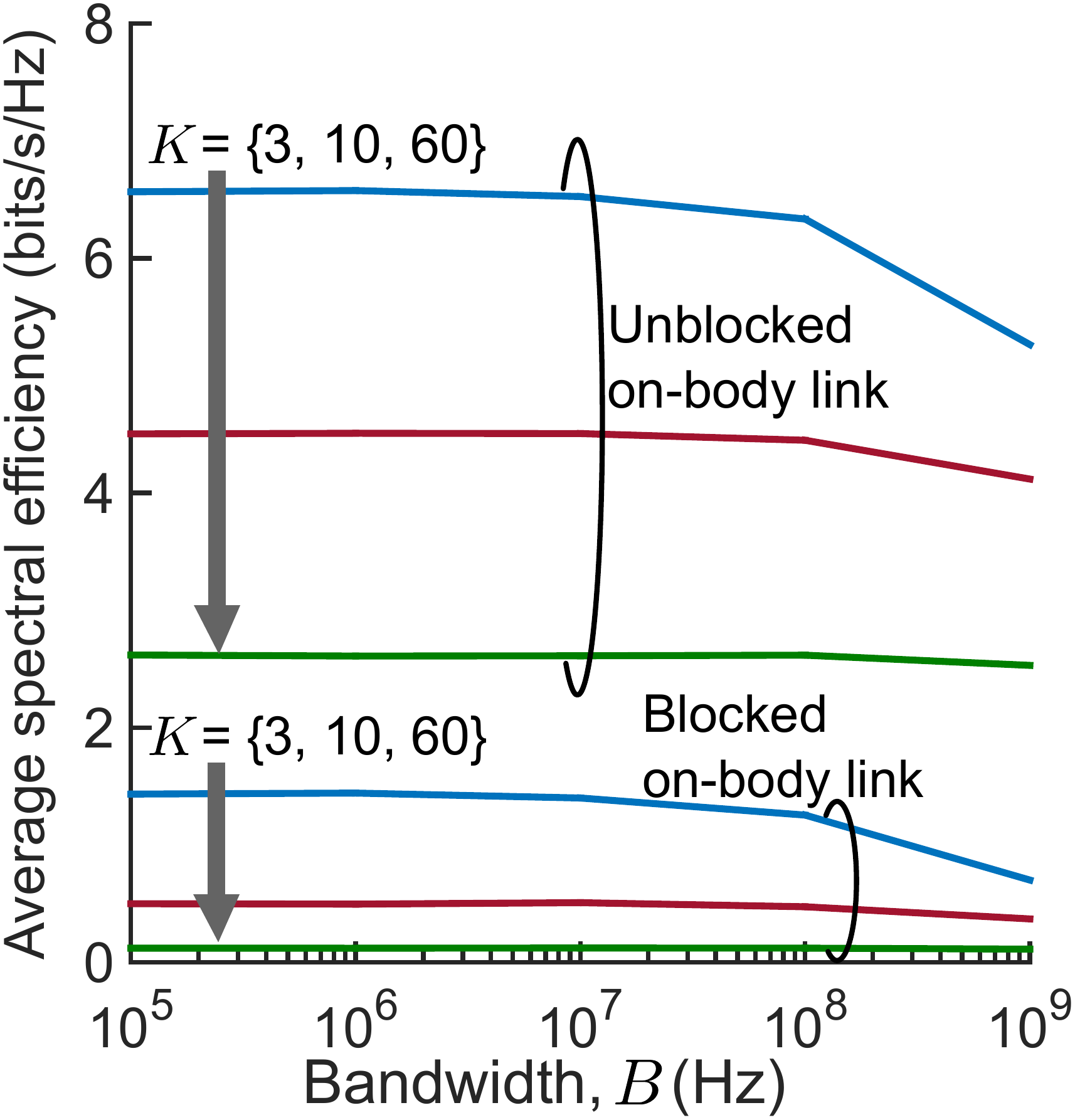}\label{Fig:figure9a}}
	\qquad
	\subfloat[Blocked on-body link ($\betaoo = 0$).]{\includegraphics[width=0.4\linewidth]{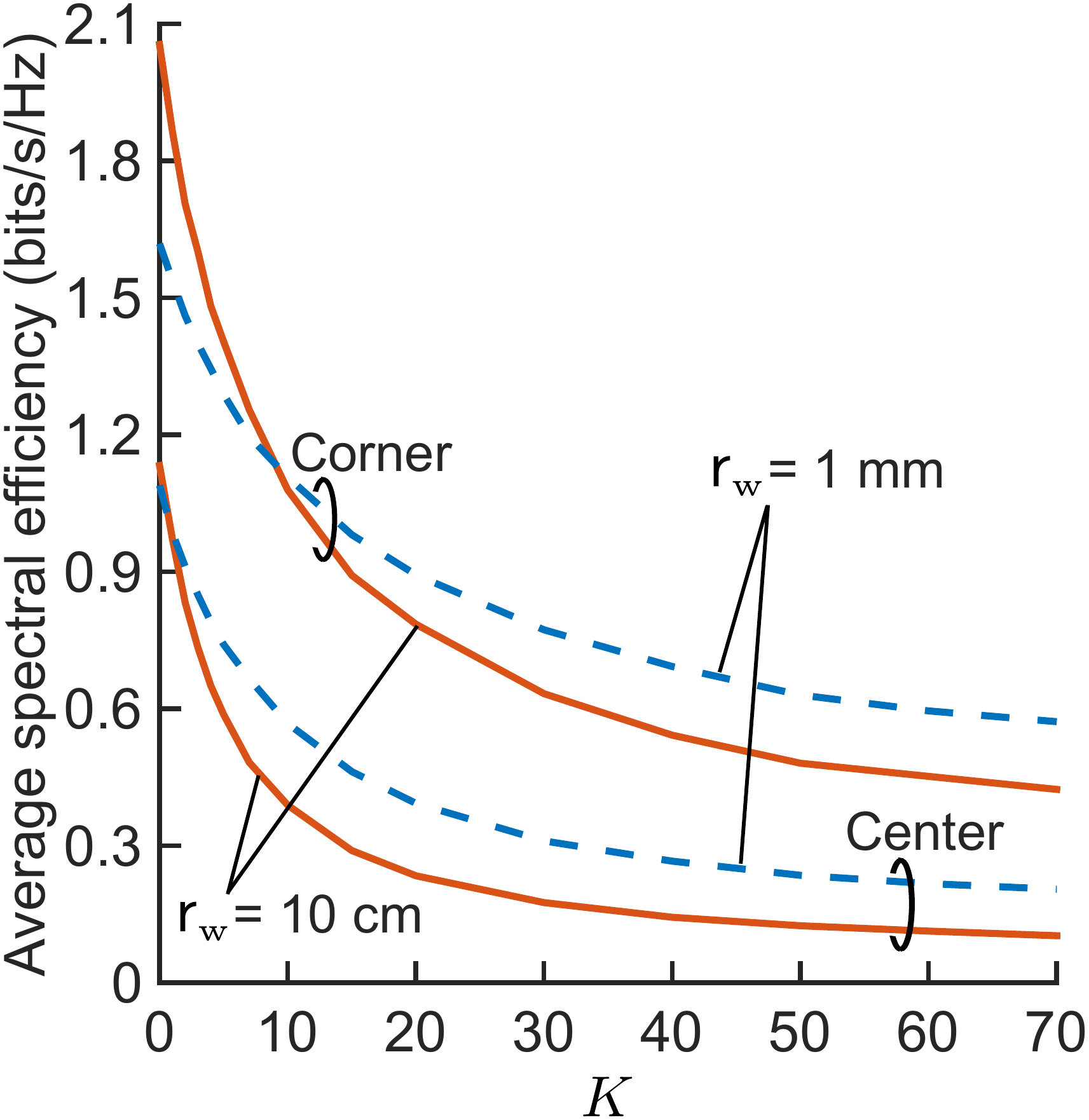}\label{Fig:figure9b}}
	\caption{Spatially averaged spectral efficiency (bits/s/Hz) with high-reflectivity surfaces ($\Gamhigh$) and isotropic antennas ($N=1$).}
	\label{Fig:figure9}
\end{figure}

\begin{exmp}
	For a reference receiver located at the center, $\bar{C}$ as function of bandwidth $B$ is plotted in Fig. \ref{Fig:figure9a}, with $\msfr = 10$ cm, high reflectivity surfaces ($\Gamhigh$) and isotropic antennas ($N = 1$). 
	Multiples curves, obtained with different values for $K$ for both unblocked on-body link ($\betaoo=1$) and blocked on-body link ($\betaoo=0$), are contrasted.
\end{exmp}
\begin{exmp}
With the bandwidth fixed again at $B=1$ GHz, Fig. \ref{Fig:figure9b} shows $\bar{C}$
as function of $K$ for high reflectivity surfaces ($\Gamhigh$), blocked on-body link ($\betaoo = 0$) and isotropic antennas ($N=1$). The contrast is between center and corner receiver locations, and between two different values of $\msfr$. 
\end{exmp}

The examples above lead to the following observations, in terms of the impact of reflections and blockages:
\begin{itemize}
\item When the direct on-body signal is unblocked, reflections are overall detrimental. The increase in interference dominates the increase in useful signal, as indicated by the degradation in performance with increasing reflectivity (cf. Fig.~\ref{Fig:figure7}).
 Given the short range of the on-body link, efficient communication is possible even without antenna gains. 
\item When the direct on-body signal is blocked and the intended signal is received only via reflections, increased reflectivity improves the performance (by as much as 10 dB), yet the SINR is very low and operation might not be feasible at all without strong antenna gains.
\item As far as interference blockages are concerned, their probability increases with the density of people and with shrinking $\msfr$, but not fast enough to fully shield receivers and hence the cumulative interference grows with the density.
\item Noise is not negligible for high-bandwidth applications (cf. Fig \ref{Fig:figure9a}), yet, even with very few interferers, the performance is on-average interference-limited. This is revealed by the steep decline in Fig. \ref{Fig:figure9b}, when going from $K=0$ (no interference) to $K>0$.
\item At corner locations, there are stronger signal reflections and a natural protection from direct interference. With very low $K$, even the effect of self-blocking in the wall-reflected signal (effected via $\msfr$) becomes noticeable, as evidenced by the cross-over of the curves in Fig. \ref{Fig:figure9b}.
\end{itemize}

Recognizing the necessity of antenna gains in the absence of a strong on-body link, the focus of the next section is on evaluating the performance improvement brought about by steering the beams towards strong signal reflections in such situations.

\section{Impact of Antenna Arrays}
\label{Imp Ant Arr}

For a reference receiver located at the center, the closest surfaces are the ceiling and the floor.
Therefore, when the on-body link is blocked, the main-lobes of the reference wearables are steered towards the signal reflection off the ceiling, which is assumed unblocked. Specifically, the main-lobe of the receiver $\Xr$ is pointed to the phantom transmitter $X_{5,0}$, while the main-lobe of the intended transmitter $\Xo$ points to the phantom image of $\Xr$ across the ceiling (cf. Appendix \ref{App1}). The azimuth and the elevation angles of these main-lobes are computed following the steps in Appendix \ref{App3}.

\begin{exmp}
\label{SE CDFs}
Plotted in Fig. \ref{Fig:SEantG} are the CDFs of spectral efficiency, $C(\mathsf{SINR})$, for a reference receiver located at the center with blocked on-body link ($\betaoo=0$), $K = 40$ and $\msfr = 10$ cm,  under different antenna settings and the two reflectivity scenarios. For $N > 1$, there are two results per case: the one in markers, obtained by applying the stochastic model for the transmit antenna parameters (cf. Section \ref{stoch mod ant G}), and the one in solid/dashed, obtained as per Appendix \ref{App3}. 
\begin{figure}[!tbp]
  \centering
  \subfloat[Low reflectivity surfaces, $\Gamlow$.]{\includegraphics[width=0.4\linewidth]{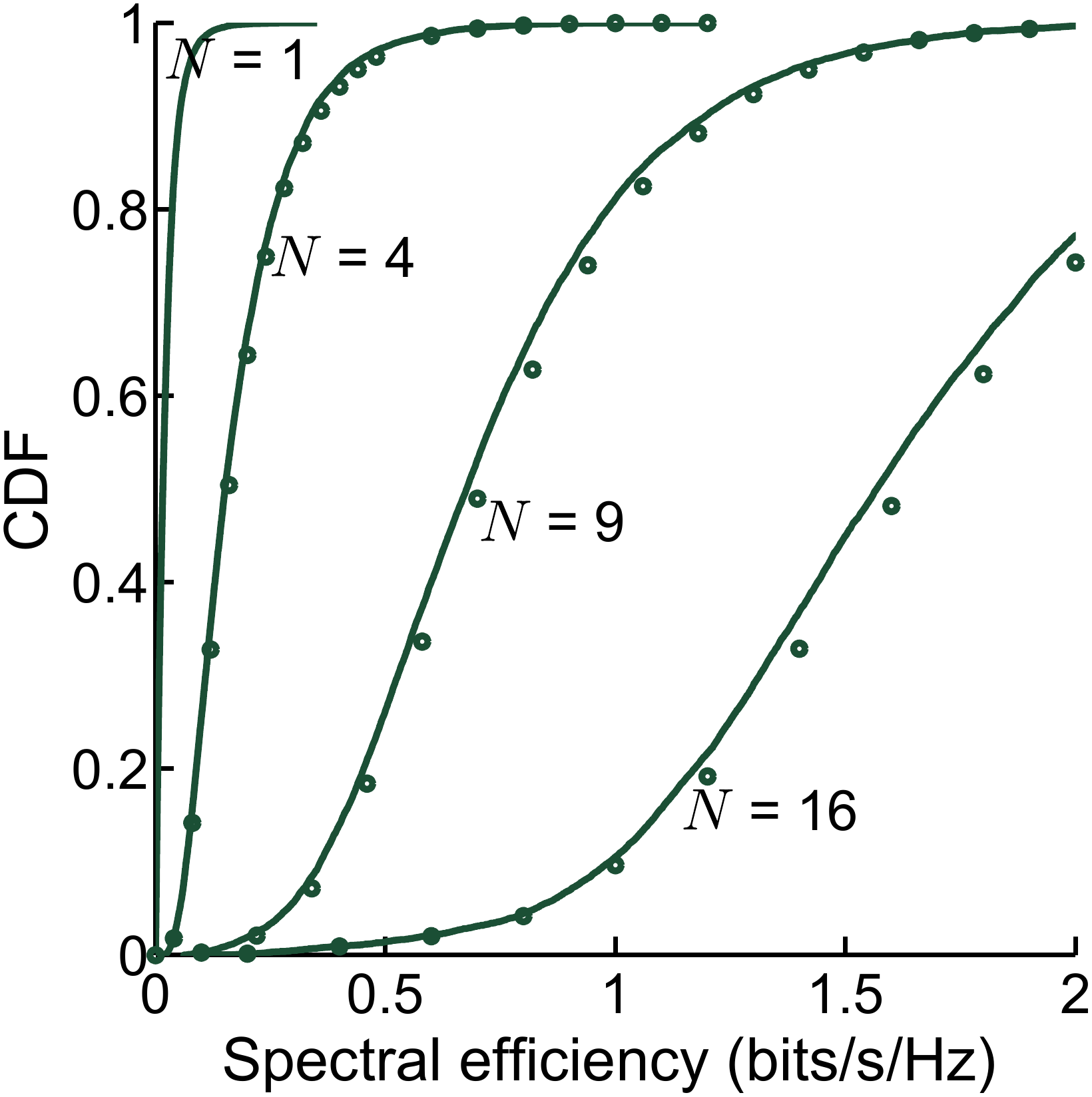}}
  \qquad
  \subfloat[High reflectivity surfaces, $\Gamhigh$.]{\includegraphics[width=0.4\linewidth]{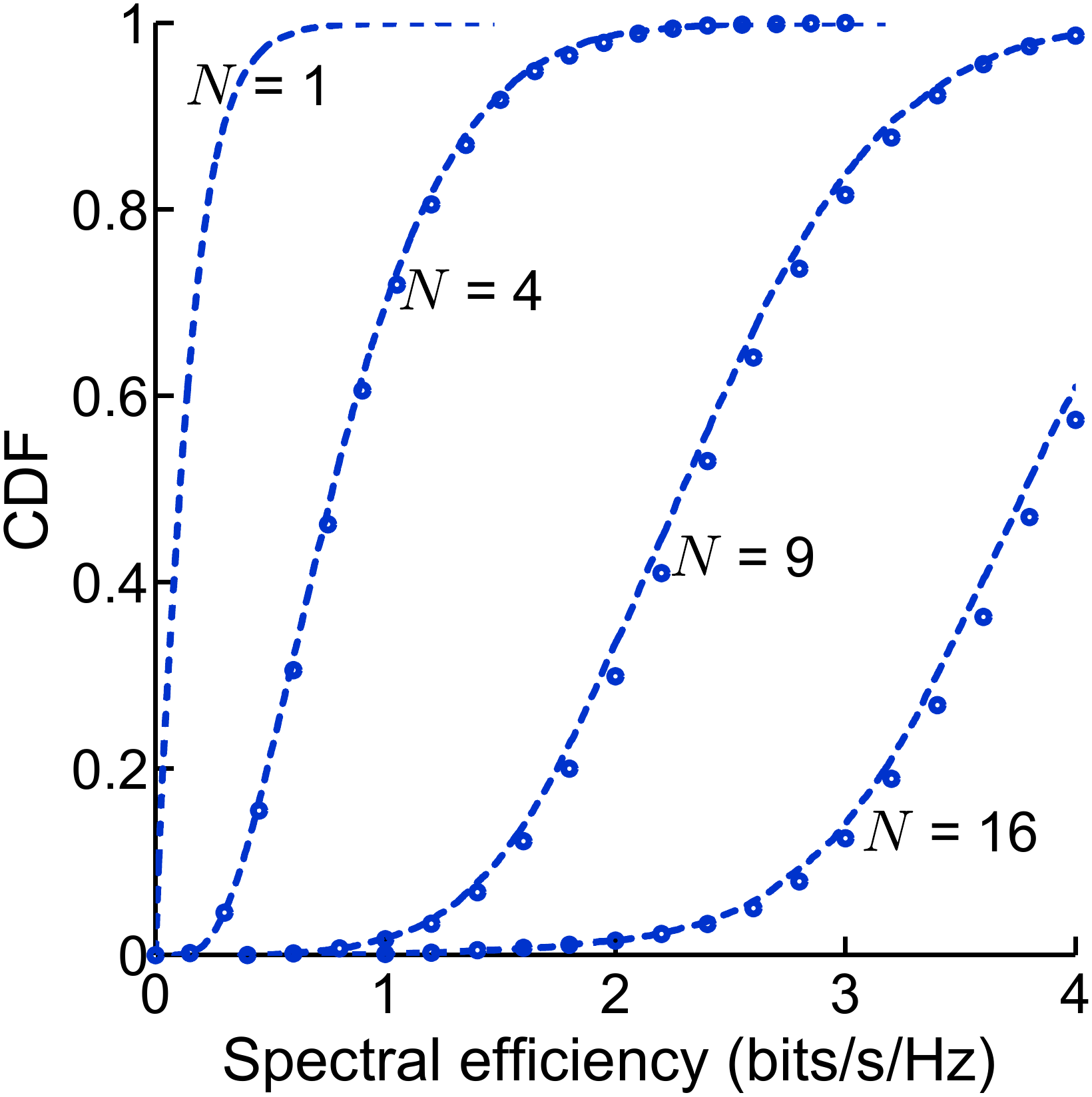}}
  \caption{CDFs of spectral efficiency (bits/s/Hz) with the reference receiver located at the center, blocked on-body link ($\betaoo = 0$), $K = 40$ and $\msfr = 10$ cm.}
  \label{Fig:SEantG}
\end{figure}
\end{exmp}
The performance improves steadily with $N$ and communication becomes feasible with high probability beyond $N \approx 9$ antennas per wearable, provided the beams are well pointed. Again, note that the dependence on surface reflectivity is significant, as much as $2$ bits/s/Hz for the settings we have considered.

Next, instead of a complete blockage of the on-body link ($\betaoo=0$), we vary $\betaoo \in (0,1]$ so as to quantify the shadow loss $1/\betaoo$ that would render reflection a better communication mechanism than the direct on-body link.
Consider two main-lobe directions for the reference wearables: (\emph{i}) towards the direct on-body link, and (\emph{ii}) towards the ceiling reflection.
We denote by $\bar{C}^{\rm o}$ and $\bar{C}^{\rm c}$ the spatially averaged spectral efficiency (\ref{eq:avgSE}) in the first and second cases, respectively.
\begin{exmp}
\label{example mean SE2}
For a reference receiver at the center,  Fig. \ref{Fig:avgSEantG} shows $\bar{C}^{\rm c} - \bar{C}^{\rm o}$ as function of $1/\betaoo$ with $\msfr=10$ cm and parameterized by $N$ and $K$.

\begin{figure}[!tbp]
  \centering
  \subfloat[Low reflectivity surfaces, $\Gamlow$.]{\includegraphics[width=0.4\linewidth]{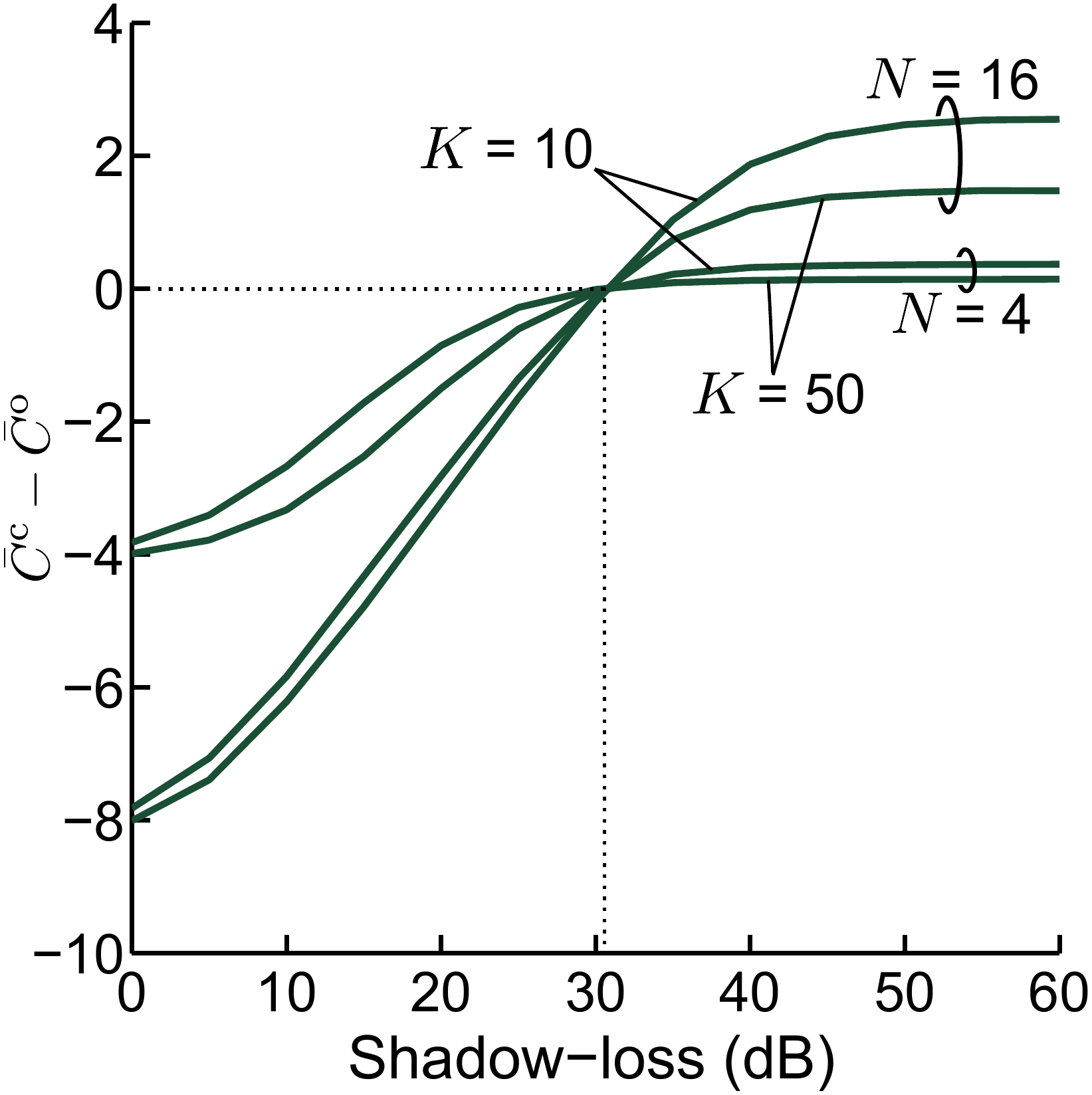}}
  \qquad
  \subfloat[High reflectivity surfaces, $\Gamhigh$.]{\includegraphics[width=0.4\linewidth]{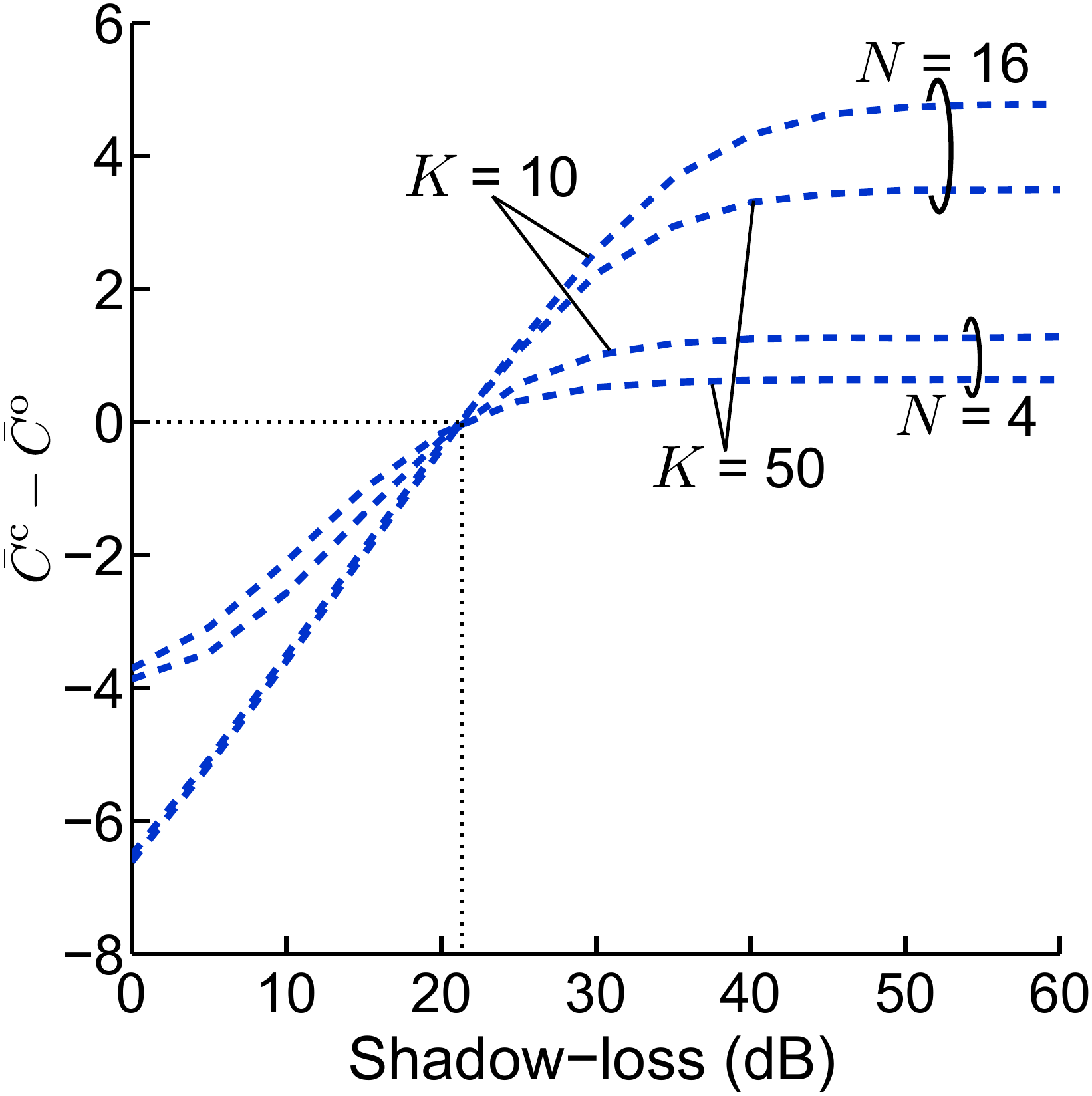}}
  \caption{Difference between the average spectral efficiencies considered in Example \ref{example mean SE2}, for varying shadow-loss in the on-body link, when the reference receiver is located at the center and $\msfr = 10$ cm.}
  \label{Fig:avgSEantG}
\end{figure}
\end{exmp}

As can be seen, steering the main-lobe to a strong signal reflection becomes preferable to the direct on-body link under relatively moderate shadowing ($20$ to $30$ dB for the settings considered), which could be rather common for on-body communication links.

Another interesting observation from Example \ref{example mean SE2} is that the shadow loss at the points where $\bar{C}^{\rm c} = \bar{C}^{\rm o}$ is roughly the same, irrespective of $N$ and $K$, for each reflectivity. This is because, in our model, the ceiling reflection of the intended signal is always available, and thus the shadow loss that renders $\bar{C}^{\rm c}$ and $\bar{C}^{\rm o}$ equal is the difference between the propagation losses of the on-body and ceiling-reflected paths.

\section{Summary}

The performance of enclosed mmWave wearable networks is influenced decidedly by blockages, reflections and directional beamforming. We have proposed a simple stochastic model that incorporates these effects, and validated the satisfactory behavior of this model under relevant settings. Further work is needed to generalize this model (e.g., to nonrectangular spaces and to include blocking/reflecting partitions within such spaces) and to extend its validation.

Indoor wearable networks may be feasible at mmWave frequencies with antenna arrays of reasonable size, even in the absence of a strong direct signal path and in high-density environments, relying on the plausible availability of signal reflections. On the order of 10 antenna elements per transceiver suffices to ensure comfortable spectral efficiencies---and thus very high bit rates given the volumes of available bandwidth---in most situations, provided the beams are well-pointed.

Potential follow-up work could include assessing the impact of body reflectivity \cite{BodyReflectionTed} (which was ignored in our models) and power control to comply with appropriate regulatory requirements for safety (e.g., the temperature-based safety compliance proposed by \cite{BodyReflectionTed}), and quantifying the degree of beam-pointing accuracy that is required, as well as devising algorithms to effect the beam pointing and tracking. Incorporation of a more refined model for the blockage/shadowing in the intended signal paths (direct on-body and ceiling/floor reflected off-body) would be key for a more comprehensive performance evaluation of wearable applications.
In scenarios that warrant inclusion of diffuse scattering effects, our model for the specular-dominant large-scale effects could be combined with stochastic small-scale fading models, similar to the quasi-deterministic approach proposed in \cite{7063558, 7499315}.

%

\section*{Acknowledgment}

The efficient editorial handling by Dr. Yindi Jing and the excellent feedback provided by the reviewers are gratefully acknowledged.

\appendices


\section{Coordinates of Image Transmitters and Angles of Incidence} \label{App1}

Recall that the origin is at the center of the enclosed space. Thus, the coordinates of the phantom transmitter $\Xik$ for $i = 1,\ldots,6$ are 
\begin{align}\nonumber
(x_{1,k}, y_{1,k}, z_{1,k}) &= (L-\xk,\yk,\zk) \\
\nonumber
(x_{2,k}, y_{2,k}, z_{2,k}) &= (-L-\xk,\yk,\zk) \\
\nonumber
(x_{3,k}, y_{3,k}, z_{3,k}) &= (\xk,W-\yk,\zk) \\
\nonumber
(x_{4,k}, y_{4,k}, z_{4,k}) &= (\xk,-W-\yk,\zk) \\
\nonumber
(x_{5,k}, y_{5,k}, z_{5,k}) &= (\xk,\yk,H-\zk) \\
\nonumber
(x_{6,k}, y_{6,k}, z_{6,k}) &= (\xk,\yk, -H-\zk)
\end{align}
and the angles of incidence are
\begin{align}\nonumber
\theta_{1,k} &= \arccos\left(|L-\xk-\xr|/r_{1,k}\right)\\
\nonumber
\theta_{2,k} &= \arccos\left(|-L-\xk-\xr|/r_{2,k}\right) \\
\nonumber
\theta_{3,k} &= \arccos\left(|W-\yk-\yr|/r_{3,k}\right) \\
\nonumber
\theta_{4,k} &= \arccos\left(|-W-\yk-\yr|/r_{4,k}\right)\\ 
\nonumber
\theta_{5,k} &= \arccos\left(|H-\zk-\zr|/r_{5,k}\right)\\
\nonumber
\theta_{6,k} &= \arccos\left(|-H-\zk-\zr|r_{6,k}\right).
\end{align}

The phantom image of the reference receiver $\Xr$ across the ceiling, mentioned in Section \ref{Imp Ant Arr}, has coordinates $(\xr, \yr, H - \zr)$.

\section{Polarization Angles}
\label{App11}

\begin{figure}
	\centering
	\includegraphics [width=0.36\columnwidth]{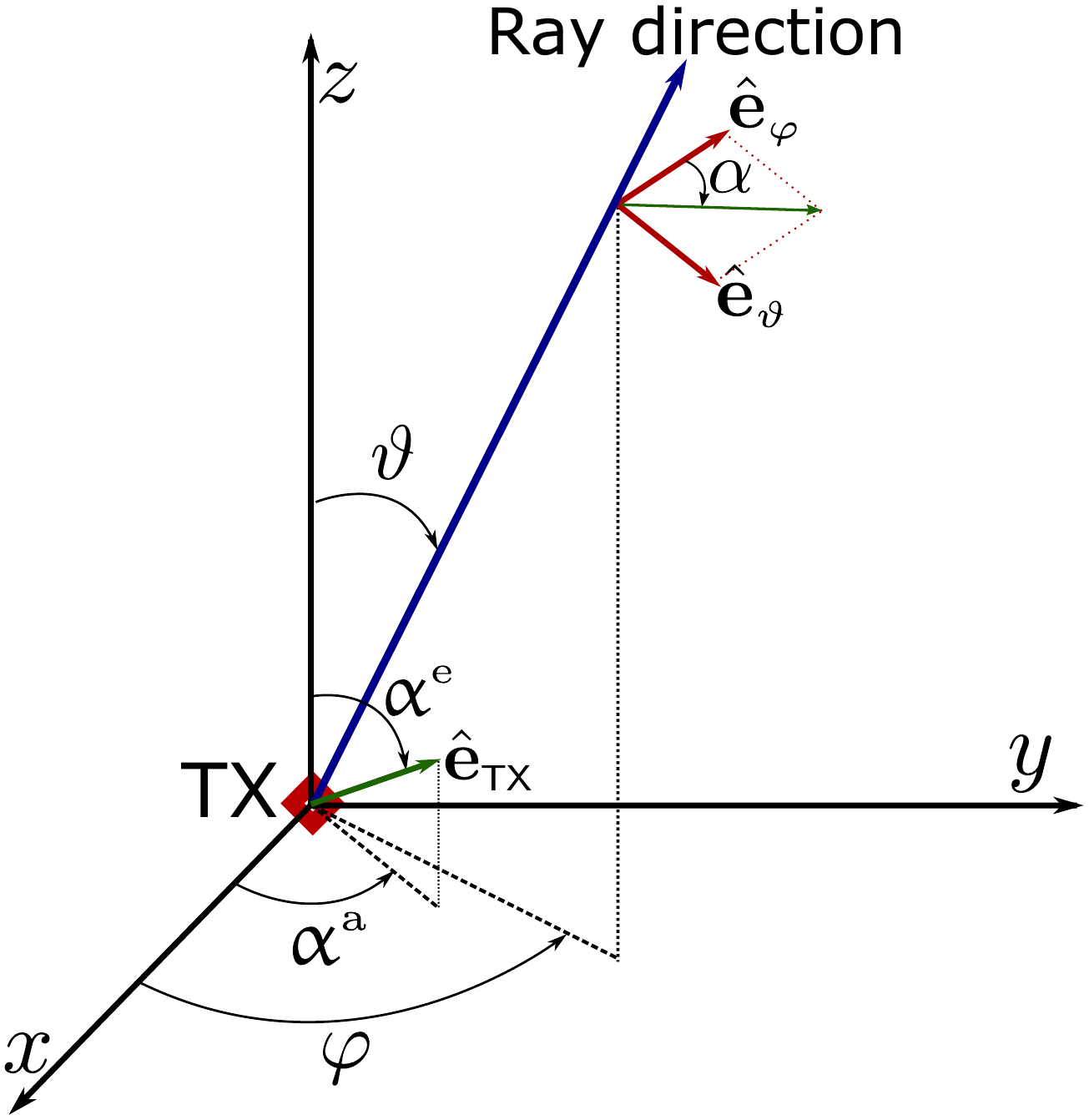}
	\caption{Polarization of a ray propagating from a transmitter (${\sf TX}$).}
	\label{Fig:pol}
\end{figure}

Consider a transmitter (cf. Fig. \ref{Fig:pol}) whose antenna polarization is defined by the azimuth angle $\upalpha^{\scriptscriptstyle \rm a}$ and the elevation angle $\upalpha^{\scriptscriptstyle \rm e}$, i.e., along the unit vector 
\begin{align}
	\hat{\bf{e}}_{\scriptscriptstyle \sf TX } &= \sin \upalpha_k^{\scriptscriptstyle \rm e} \, \cos \upalpha_k^{\scriptscriptstyle \rm a} \, \hat{\pmb{x}} + \sin \upalpha_k^{\scriptscriptstyle \rm e} \, \sin \upalpha_k^{\scriptscriptstyle \rm a} \, \hat{\pmb{y}} + \cos \upalpha_k^{\scriptscriptstyle \rm e} \, \hat{\pmb{z}}.
\end{align}
Then, the polarization in a ray propagating from the transmitter in a direction specified by azimuth $\varphi$ and elevation $\vartheta$ is abstracted by the angle
\begin{align}\label{eq:pol}
  \alpha(\vartheta,\varphi) = \arctan\left(\frac{|\langle\hat{\bf{e}}_{\scriptscriptstyle \sf TX }, \hat{\bf{e}}_{\scriptscriptstyle \vartheta} \rangle|}{|\langle\hat{\bf{e}}_{\scriptscriptstyle \sf TX }, \hat{\bf{e}}_{\scriptscriptstyle \varphi} \rangle|}\right)
\end{align}
where $\langle\cdot,\cdot\rangle$ denotes the inner product and the unit normal vectors perpendicular to the ray direction (cf. Fig. \ref{Fig:pol}) are
\begin{align}
\hat{\bf{e}}_{\scriptscriptstyle \vartheta} &= \cos \vartheta \, \cos \varphi \, \hat{\pmb{x}} + \cos \vartheta \, \sin \varphi \, \hat{\pmb{y}} - \sin \vartheta \, \hat{\pmb{z}} \\
\hat{\bf{e}}_{\scriptscriptstyle \varphi} &= \sin \varphi \, \hat{\pmb{x}} - \cos \varphi \, \hat{\pmb{y}}.
\end{align}
Namely, the electric field vector is in the direction of $\hat{\bf{e}}_{\scriptscriptstyle \varphi} \, \cos [\alpha(\vartheta,\varphi)] + \hat{\bf{e}}_{\scriptscriptstyle \vartheta} \, \sin [\alpha(\vartheta,\varphi)]$.

Let the orientation of the transmitter location $\Xk$ with respect to the reference receiver $\Xr$ be specified by the elevation and azimuth angles
\begin{align}\label{eq:momu}
\txorientEk &= \arccos \frac{\zk-\zr}{\rk} \\
\label{eq:momu1}
\txorientAk 
&= \arg [(\xk-\xr) + \imunit \, (\yk-\yr)].
\end{align}
Similarly, let the orientation of the image location $\Xik$ with respect to $\Xr$ be
\begin{align}
\label{eq:juggu}
\txorientEik &= \arccos \frac{\zik-\zr}{\rik} \\
\label{eq:juggu1}
\txorientAik 
&= \arg [(\xik-\xr) + \imunit \,  (\yik-\yr)].
\end{align}

We model the antenna polarization of the transmitter $X_k$ by a uniformly distributed azimuth angle $\upalpha_k^{\scriptscriptstyle \rm a} \in [0,2\pi)$, and an elevation angle $\upalpha_k^{\scriptscriptstyle \rm e}$ distributed with PDF $f_{\upalpha_k^{\scriptscriptstyle \rm e}}(\nu) = \frac{\sin \nu}{2} \, 0 \leq \nu \leq \pi.$
Then, following (\ref{eq:pol}), the polarization angles (in each propagation path) are computed as
\begin{align}
\alpha_k &= \arctan \left( \frac{|\cos(\txorientEk)\,\cos(\upalpha_k^{\scriptscriptstyle \rm a}-\txorientAk) \, \sin(\upalpha_k^{\scriptscriptstyle \rm e}) - \cos(\upalpha_k^{\scriptscriptstyle \rm a})\,\sin(\txorientEk)|}{|\sin(\upalpha_k^{\scriptscriptstyle \rm e}) \, \sin(\upalpha_k^{\scriptscriptstyle \rm a} - \txorientAk)|}\right)\\
\alpha_{i,k} &= \arctan \left( \frac{|\cos(\txorientEik)\,\cos(\upalpha_k^{\scriptscriptstyle \rm a}-\txorientAik) \, \sin(\upalpha_k^{\scriptscriptstyle \rm e}) - \cos(\upalpha_k^{\scriptscriptstyle \rm a})\,\sin(\txorientEik)|}{|\sin(\upalpha_k^{\scriptscriptstyle \rm e}) \, \sin(\upalpha_k^{\scriptscriptstyle \rm a} - \txorientAik)|}\right).
\end{align}




\section{Algorithm to Determine Blockages}\label{App2}

\subsection{Direct Interference Paths and Wall Reflections}
\label{App2 sec1}

This is a modified version of the algorithm presented in \cite{KirHeath15, mmWaveD2D_Kiran15} that includes blocking of the wall reflections.
Let the circles (people), and more precisely the locations of their centers on $\mathX$, be denoted by $\{\Yk\}_{k=0}^K$. The phantom images of $\Yk$,  across the four walls, are denoted by $\{\Yik\}_{i=1}^4$.
Given $\Xr$, $\{\Xk\}$ and $\{\Yk\}$, the blockages in the direct interfering paths and the paths reflected off the walls can be determined using the following algorithm:
\begin{enumerate}
\item Let $\mathcal{S}_{X^{\scriptscriptstyle \rm c}} = \{\Yk\} \cup \{\Yik\}_{i=1}^4$ be the set of circles on the horizontal plane $\mathX$.
For each $X^{\scriptscriptstyle \rm c}_\ell \in \mathcal{S}_{X^{\scriptscriptstyle \rm c}}$, compute the distance $\|X^{\scriptscriptstyle \rm c}_\ell - \Xr\|$ and the corresponding angle $\angle (X^{\scriptscriptstyle \rm c}_\ell - \Xr)$ 
\item Compute the blocking cones formed by each circle $X^{\scriptscriptstyle \rm c}_\ell$ as \cite{KirHeath15, mmWaveD2D_Kiran15}
\begin{align}
B_{{\rm \scriptscriptstyle C}_{\scriptscriptstyle \ell}} &= \angle (X^{\scriptscriptstyle \rm c}_\ell - \Xr) \pm \arcsin\frac{\usrDia}{2 \, \|X^{\scriptscriptstyle \rm c}_\ell - \Xr\|}
\end{align}
\item Determine $\mathcal{S}_{X} = \{\Xk\} \cup \{\Xik\}_{i=1}^4$, the set of all the transmitters and their phantom images across the walls.
\item A transmitter $X_m \in \mathcal{S}_{X}$ is blocked when either of the following two conditions are true:
\begin{itemize}
\item $X'_m$, the projection of $X_m$ on $\mathX$, lies within the blockage cones of the circles with $\|X^{\scriptscriptstyle \rm c}_\ell - \Xr\| < \|X'_m - \Xr\|$, i.e.,
\begin{align}
\angle (X'_m - \Xr) \in \bigcup_{\{\ell: \, \|X^{\scriptscriptstyle \rm c}_\ell-\Xr\| < \|X'_m-\Xr\|\}} B_{{\rm \scriptscriptstyle C}_{\scriptscriptstyle \ell}}
\end{align}
\item $X'_m$ has any $X^{\scriptscriptstyle \rm c}_\ell \in \mathcal{S}_{X^{\scriptscriptstyle \rm c}}$ within a distance $\usrDia/2$.
\end{itemize}
\end{enumerate}

\subsection{Ceiling Reflected Interference Path}
\label{App2 sec2}

\begin{figure}
	\centering
	\includegraphics [width=0.5\columnwidth]{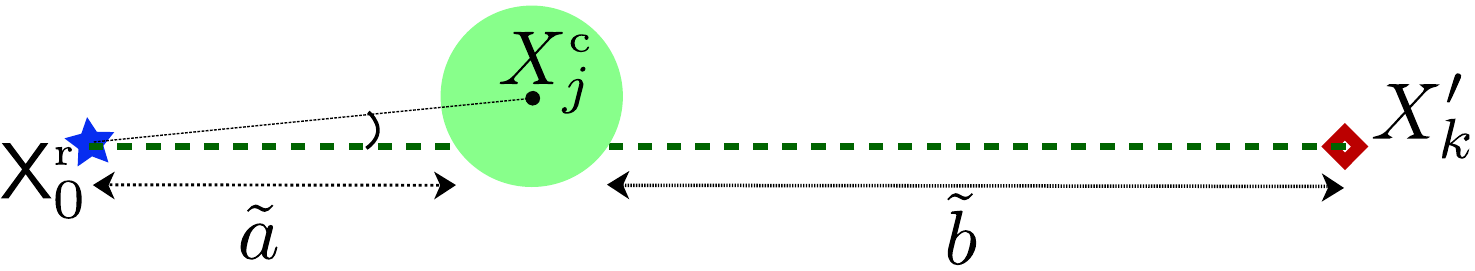}
	\caption{A blockage ($X^{\scriptscriptstyle \rm c}_j$) in the direct path between $\Xk'$ and $\Xr$.}
	\label{Fig:ceilingref}
\end{figure}

For each blockage in the direct path with circle center $X^{\scriptscriptstyle \rm c}_j$, compute (cf. Fig. \ref{Fig:ceilingref})
\begin{align}
\aktilde &= d \cos \xi - (\usrDia/2) \cos \! \big( \! \arcsin[(2 d / \usrDia) \sin \xi] \big) \\
\bktilde &= \rk' - d \cos \xi - (\usrDia/2) \cos \! \big( \!  \arcsin[(2 d / \usrDia) \sin \xi] \big)
\end{align}
where $d = \|X^{\scriptscriptstyle \rm c}_j - \Xr\|$ and $\xi = \angle (X^{\scriptscriptstyle \rm c}_j - \Xr)$.
The ceiling reflection from $\Xk$ is blocked ($\beta_{5,k}=0$) by the user circle $X^{\scriptscriptstyle \rm c}_j$ if $\tilde{a} < a_k$ or $\tilde{b} < b_k$.

\section{Antenna Gains}
\label{App3}

Recall the orientations given in (\ref{eq:momu}), (\ref{eq:momu1}), (\ref{eq:juggu}) and (\ref{eq:juggu1}).
Then, the receive antenna gains are obtained via (\ref{eq:antpat}) as
\begin{align}
 \Grk &= \patG(\vartheta)|_{\vartheta = \arccos[\cos\RXbeamE \, \cos\txorientEk + \sin\RXbeamE \, \sin\txorientEk \, \cos(\txorientAk - \RXbeamA)]} \\
 \Grik &= \patG(\vartheta)|_{\vartheta = \arccos[\cos\RXbeamE \, \cos\txorientEik + \sin\RXbeamE \, \sin\txorientEik \, \cos(\txorientAik - \RXbeamA)]}
\end{align}
and the transmit antenna gains become 
\begin{align}
 \Gtk &= \patG(\vartheta)|_{\vartheta = \arccos[-\cos\TXbeamEk \, \cos\txorientEk - \sin\TXbeamEk \, \sin\txorientEk \, \cos(\txorientAk - \TXbeamAk)]} \\
 \Gtik &= \patG(\vartheta)|_{\vartheta = \arccos[- \cos\TXbeamEk \, \cos  \txorientEik - \sin\TXbeamEk \, \sin \txorientEik \, \cos( \txorientAik - \TXbeamAk)]} 
\end{align}
%
where 
\begin{align}\nonumber
\TXbeamEik &= \TXbeamEk \quad i = 1,\ldots,4  \\
\nonumber
\TXbeamEik &= \pi - \TXbeamEk \quad i = 5,6 \\
\TXbeamAik &= \pi - \TXbeamAk \quad i = 1,2\\
\nonumber
\TXbeamAik &= - \TXbeamAk \quad i = 3,4\\
\nonumber
\TXbeamAik &=  \TXbeamAk \quad i = 5,6.
\end{align}
specify the main-lobe directions of the phantom transmitters.

\section{Derivation of Blockage Probability}\label{App4}

\subsection{Direct Interference Paths}
\begin{figure}
	\centering
	\includegraphics [width=0.5\columnwidth]{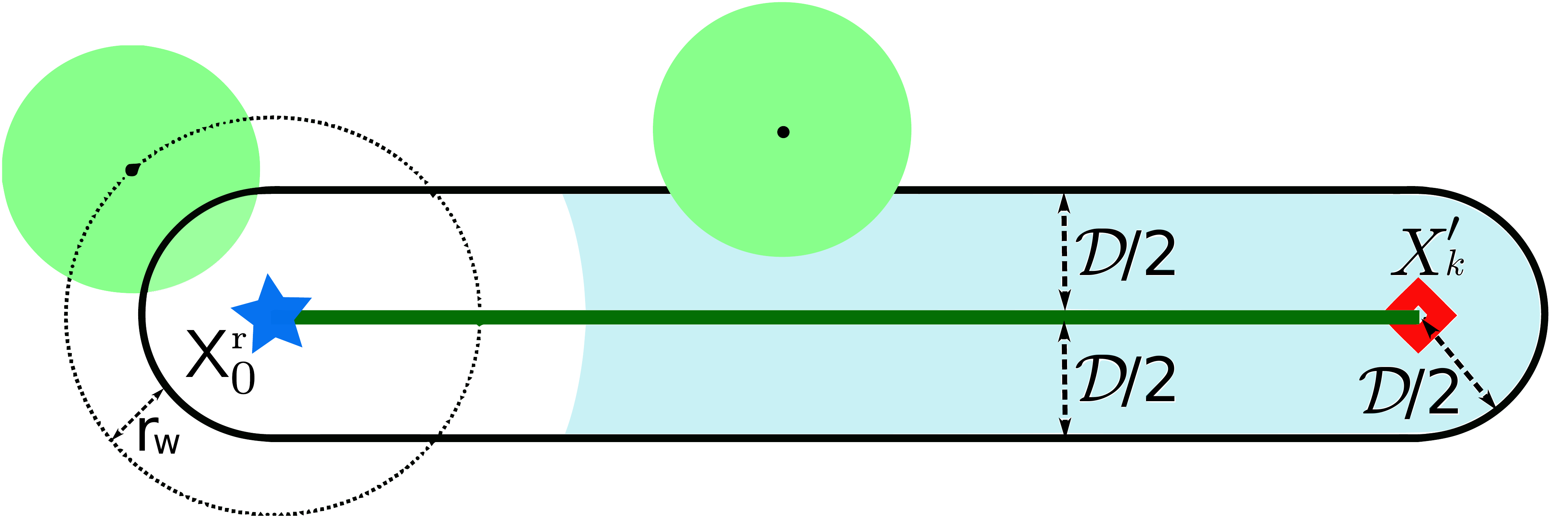}
	\caption{A direct interferer path. The link intersects any circle with center inside the 2-D capsule.}
	\label{Fig:figure3}
\end{figure}
Consider the projection of a direct interference link from $\Xk$ to $\Xr$ on $\mathX$, as depicted in Fig. \ref{Fig:figure3}.  A circle of diameter $\usrDia$ is intersected by the link if and only if its center falls inside the 2-D capsule of radius $\usrDia/2$ and length $\rk' = \|\Xk'-\Xr\|$ drawn around the link. The circle centers, $\Yo$ and $\Yk$, each fall independently inside the capsule (self-body blocking) with probability 
\be\label{eq:selfblk}
p_{\rm \scriptscriptstyle sb} = \frac{\arcsin\frac{\usrDia}{2 \, \msfr+\usrDia}}{\pi}.
\ee
Each of the $K-1$ other circle centers in $\{\Yj\}_{j \notin \{0,k\}}$ is located at a distance $\usrDia/2+\msfr$ away from the corresponding wearable $X_j$, at an angle $\varrho_j$ independently and uniformly distributed in $[0,2\pi)$, with coordinates $\xcj = x_j + (\usrDia/2+\msfr) \cos \varrho_j$ and $\ycj = y_j + (\usrDia/2+\msfr) \sin \varrho_j$.
Then, $\{\Yj\}_{j \notin \{0,k\}}$ are independently distributed on $\mathX$, with PDF
\begin{align}
f_{\xcj,\ycj}(x,y) &= \int_{0}^{2\pi}f_{\xk,\yk}\left[x-(\usrDia/2+\msfr) \cos \varrho ,y - (\usrDia/2+\msfr) \sin \varrho\right] {\rm d}\varrho
\end{align}
which we approximate as
\begin{align}\label{eq:uniform circles}
f_{\xcj,\ycj}(x,y) &\approx f_{\xk,\yk}(x,y)
\end{align}
the accuracy of which, in modeling the quantities of interest, is validated via Examples \ref{av block prob} and \ref{SINR CDF}. 

Each circle center in $\{\Yj\}_{j \notin \{0,k\}}$ independently fall inside the capsule with probability
\be\label{eq:normblk}
p_{\rm \scriptscriptstyle ob} (\rk') \approx \frac{\rk' \, \usrDia - \mathcal{A} }{LW - \pi (\usrDia+\msfr)^2}
\ee
where $\mathcal{A} + \pi \usrDia^2/8$ is the area of the intersection of the capsule and the exclusion circle, i.e., the unshaded part of the capsule in Fig. \ref{Fig:figure3}. 
Thus, the probability of the link being blocked satisfies
\begin{align}\label{eq:prob2}
\mathbb{P}[\betaok = 0]  &\approx 1 - \left( 1 - p_{\rm \scriptscriptstyle ob} (\rk')\right)^{K-1} \left(1 - p_{\rm \scriptscriptstyle sb} \right)^2.
\end{align}
The result in (\ref{eq:probblk}) is obtained by plugging (\ref{eq:selfblk}) and (\ref{eq:normblk}) into (\ref{eq:prob2}).

\subsection{Wall-Reflected Interference Paths}

\begin{figure}
	\centering
	\includegraphics [width=0.45\columnwidth]{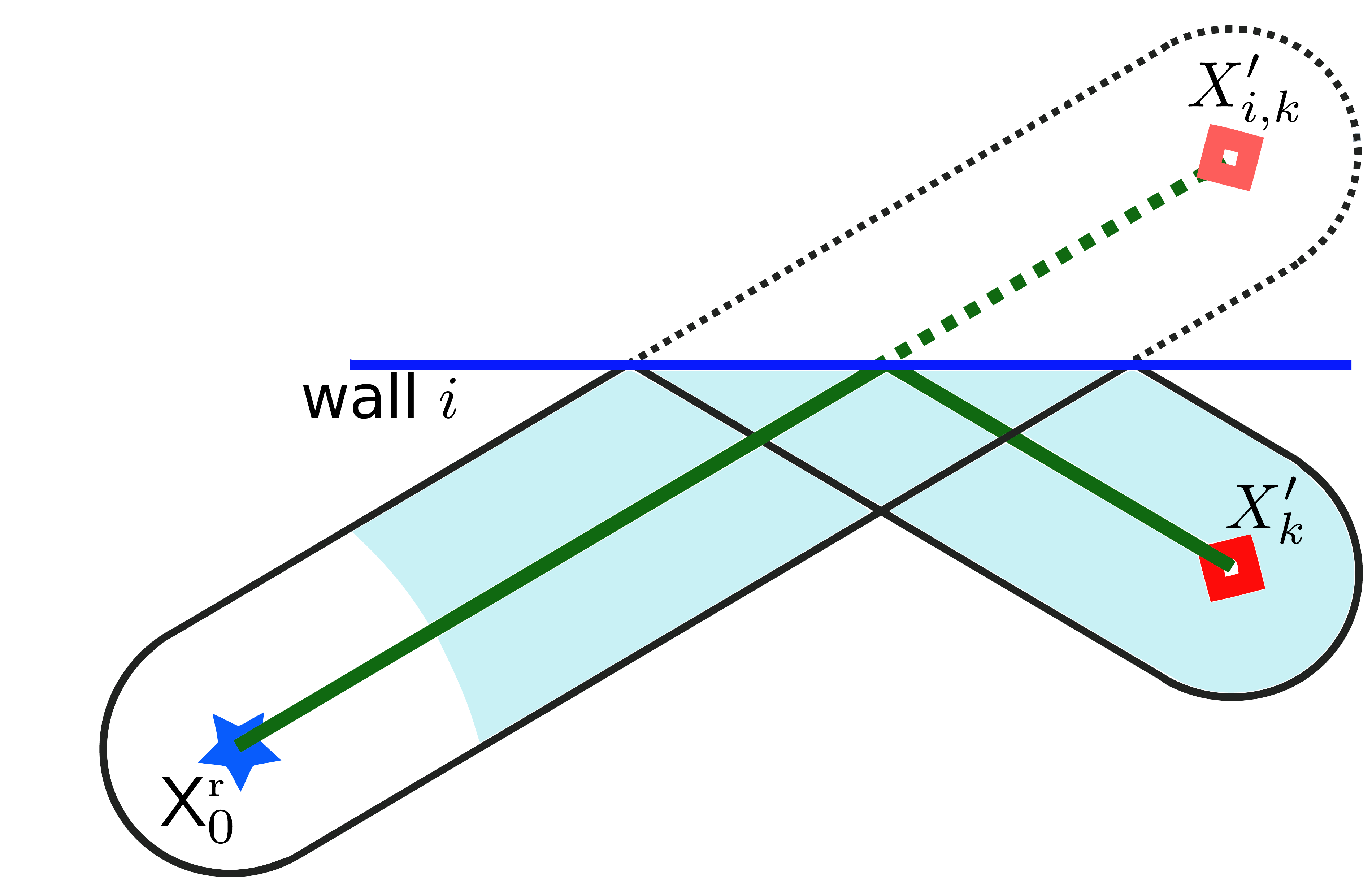}
	\caption{A reflected interferer link off a wall. The link intersects any circle with center inside the folded 2-D capsule depicted with solid lines.}
	\label{Fig:figure5}
\end{figure}

The link from a phantom transmitter across the walls, $\{\Xik\}_{i=1}^4$, is blocked by circles with center falling inside the folded 2-D capsule in Fig. \ref{Fig:figure5}.
Then, (\ref{eq:prob2}) can be used as a close approximation for the blockage probability of the wall reflections, not exact only because of the folded capsule having slightly lesser area than the unfolded capsule (cf. Fig. \ref{Fig:figure5}).

\subsection{Ceiling-Reflected Interference Paths}
\begin{figure}
	\centering
	\includegraphics [width=0.45\columnwidth]{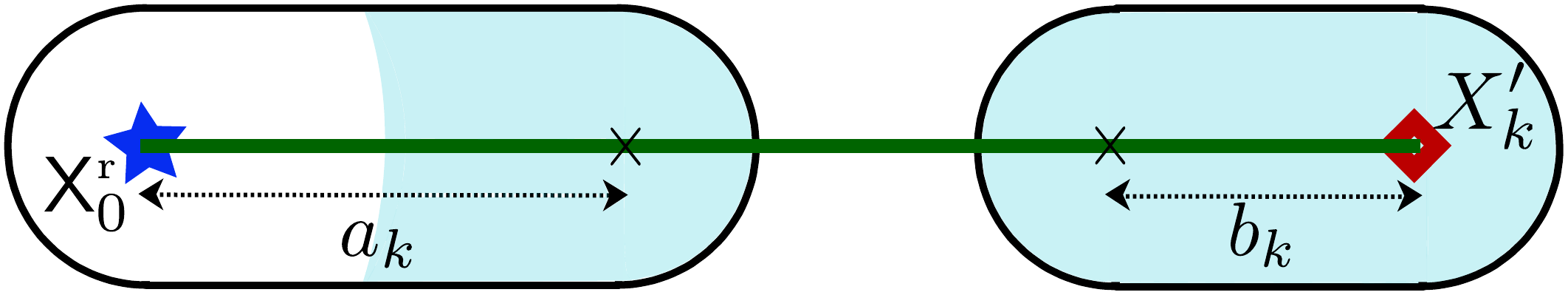}
	\caption{A direct interferer link. The corresponding ceiling reflection intersects any circle with center inside any of the two 2-D capsules.}
	\label{Fig:celingproof}
\end{figure}

Those blockages in the direct path from $\Xk$ that block the ceiling reflection as per Section \ref{blockage interference} will have their circle centers inside one of the two 2-D capsules depicted in Fig. \ref{Fig:celingproof}. Recall from Section \ref{blockage interference} that $\ak = (\usrHt - H/2 - \zr) \, \tan \theta_{5,k}$ and $\bk = (\usrHt - H/2 - \zk) \, \tan \theta_{5,k}$ depend on the wearable heights. 
Consider the self-body blocking by the reference user $\Yo$, which is located $\usrDia/2 +\msfr$ away from $\Xr$ and at a uniformly random angle in $[0,2\pi)$. The probability for $\Yo$ falling in the capsule of length $\ak$ becomes the self-body blocking probability in (\ref{eq:celilSB0}). Similarly, $\Yk$, located $\usrDia/2 +\msfr$ away from $\Xk$ and at a uniformly random angle in $[0,2\pi)$, effects self-body blocking with probability (\ref{eq:celilSBk}).
Again, the probability of any of the other $K-1$ users falling in either of the two capsules, given by (\ref{eq:eq:celilOB}), is obtained via the density in (\ref{eq:uniform circles}). 

To capture the dependence between $\beta_{5,k}$ and $\betaok$, we introduce an auxiliary random variable $\tilde{\beta}_k$, which is Bernoulli with
\begin{align}\label{eq:betatildek}
\mathbb{P}[\tilde{\beta}_k = 0] = \mathbb{P}[\betaok = 0 | \beta_{5,k} = 1] .
\end{align}
This is essentially the probability of the direct link being blocked, given that no circle center is present inside the two capsules in Fig. \ref{Fig:celingproof}.
Then, $\betaok$ computed as
\begin{align}
\label{eq:betaok}
\betaok =  \tilde{\beta}_{k} \, \beta_{5,k}
\end{align}
by independently generating $\beta_{5,k}$ and $\tilde{\beta}_{k}$, satisfies
\begin{align}
\mathbb{P}[\betaok = 0] &= \mathbb{P}[\tilde{\beta}_k = 0] \, \mathbb{P}[\beta_{5,k} = 0] + \mathbb{P}[\tilde{\beta}_k = 1] \, \mathbb{P}[\beta_{5,k} = 0] + \mathbb{P}[\tilde{\beta}_k = 0] \, \mathbb{P}[\beta_{5,k} = 1] \\
&=  \mathbb{P}[\beta_{5,k} = 0] + \mathbb{P}[\tilde{\beta}_k = 0] \, \mathbb{P}[\beta_{5,k} = 1]
\end{align}
complying with (\ref{eq:directblockprob}).
Explicitly, $\tilde{\beta}_{k}$ as per (\ref{eq:betatildek}) can be modeled as
\be
\tilde{\beta}_k = \tilde{\upbeta}_{k}^{\scriptscriptstyle {\rm sb}0} \, \tilde{\upbeta}_{k}^{\scriptscriptstyle {\rm sb}k} \, \tilde{\upbeta}_{k}^{\scriptscriptstyle \rm ob}
\ee
where the factors are independent Bernoulli random variables with probabilities
\begin{align} 
\mathbb{P}[\tilde{\upbeta}_{k}^{\scriptscriptstyle {\rm sb}0} = 0] = \left\{ \begin{array}{l l}
 0 \quad &  \quad \ak \geq \sqrt{\msfr (\usrDia+\msfr)} \\
  \frac{ \arcsin\frac{\usrDia}{2 \, \msfr+\usrDia} - \arccos\frac{\ak^2 +  \msfr \usrDia + \msfr^2}{\ak (2 \, \msfr+\usrDia)}}{\pi - \arccos\frac{\ak^2 +  \msfr \usrDia + \msfr^2}{\ak (2 \, \msfr+\usrDia)}}  \quad &  \quad \sqrt{\msfr (\usrDia+\msfr)} > \ak \geq \msfr \\ 
\frac{\arcsin\frac{\usrDia}{2 \, \msfr+\usrDia}}{\pi} \quad &  \quad \ak < \msfr
 \end{array} \right.
\end{align}
\begin{align} 
\mathbb{P}[\tilde{\upbeta}_{k}^{\scriptscriptstyle {\rm sb}k} = 0] = \left\{ \begin{array}{l l}
 0 \quad &  \quad \bk \geq \sqrt{\msfr (\usrDia+\msfr)} \\
  \frac{ \arcsin\frac{\usrDia}{2 \, \msfr+\usrDia} - \arccos\frac{\bk^2 +  \msfr \usrDia + \msfr^2}{\bk (2 \, \msfr+\usrDia)}}{\pi - \arccos\frac{\bk^2 +  \msfr \usrDia + \msfr^2}{\bk (2 \, \msfr+\usrDia)}}  \quad &  \quad \sqrt{\msfr (\usrDia+\msfr)} > \bk \geq \msfr \\ 
\frac{\arcsin\frac{\usrDia}{2 \, \msfr+\usrDia}}{\pi} \quad &  \quad \bk < \msfr
 \end{array} \right.
\end{align}
\begin{align}
\mathbb{P}[\tilde{\upbeta}_k^{\scriptscriptstyle \rm ob} = 0] 
&\approx 1 - \left( \frac{\mathsf{Area} - \rik'  \usrDia + \mathcal{A}  }{\mathsf{Area} -  (\ak + \bk) \, \usrDia + \mathcal{A} - \pi \, \usrDia^2/2}\right)^{K-1}
\end{align} 
where $\mathsf{Area} = LW - \pi (\usrDia+\msfr)^2$.

\subsection{Wall-Reflected Signal Path}
   \begin{figure}
 	\centering
 	\includegraphics [width=0.6\columnwidth]{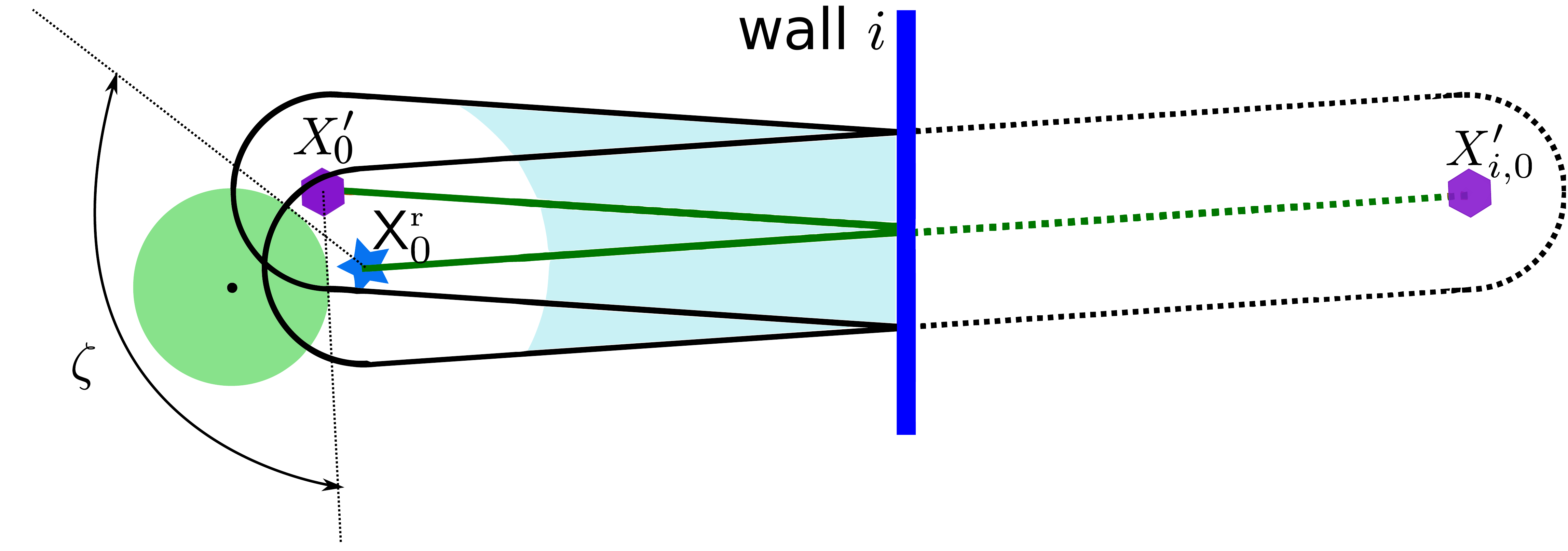}
 	\caption{A reflected signal link off a wall. The link intersects any circle with center inside the folded  2-D capsule depicted with solid lines.}
 	\label{Fig:figure8}
 \end{figure}
For the signal reflections off the walls, i.e., the links from $\{\Xio\}_{i=1}^4$, only the reference user $\Yo$ can cause self-body blocking and the other $K$ users can potentially block the link if any circle center falls within the folded capsule depicted in Fig. \ref{Fig:figure8}. 
Specifically, self-body blocking of the $i$th reflection happens when it falls in the angle $\zeta$ depicted in Fig. \ref{Fig:figure8} and computed as
\be
\zeta = 2\left(\arcsin\frac{\ro'}{2 \, \msfr+\usrDia} + \arcsin\frac{\usrDia}{2 \, \msfr+\usrDia}\right).
\ee
Then, we approximate the folded capsule area with half the area of the unfolded capsule (cf. Fig. \ref{Fig:figure8}) and
an approximation of the blockage probability of the reflected signal link off the $i$th wall can be obtained as in (\ref{eq:block prob sig ref}).

\bibliographystyle{IEEEtran}
\bibliography{jour_short,conf_short,references}
\end{document}